# Local complexity predicts global synchronization of hierarchically networked oscillators

Jin Xu,[1,2,a)] Dong-Ho Park,[1,a)] and Junghyo Jo[1,2,b)]

[1]*Asia Pacific Center for Theoretical Physics (APCTP), 67 Cheongam-ro, Pohang, 37673, Korea*

[2]*Department of Physics, Pohang University of Science and Technology (POSTECH), Pohang, 37673, Korea*

We study the global synchronization of hierarchically-organized Stuart-Landau oscillators, where each subsystem consists of three oscillators with activity-dependent couplings. We consider all possible coupling signs between the three oscillators, and find that they can generate different numbers of phase attractors depending on the network motif. Here, the subsystems are coupled through mean activities of total oscillators. Under weak inter-subsystem couplings, we demonstrate that the synchronization between subsystems is highly correlated with the number of attractors in uncoupled subsystems. Among the network motifs, perfect anti-symmetric ones are unique to generate both single and multiple attractors depending on the activities of oscillators. The flexible local complexity can make global synchronization controllable.

**Biological and social systems have hierarchical organization with heterogeneous elements. It is challenging to understand the synchronization behavior of such complex systems. In this study, we demonstrate a simple hypothesis that the local complexity of subsystems can govern their global synchronization. In particular, anti-symmetric couplings between subsystem elements are effective to alter their local complexity. This suggests that the ubiquitous appearance of asymmetric couplings can be an effective strategy to control global synchronization in real-world networks.**

## I. INTRODUCTION

Synchronization is a representative phenomenon of coherence in living systems.[1-3] Rhythms and synchronization of neural activities are necessary for odor discrimination,[4] visual feature integration,[5] and brain computation,[6] but perfect synchronization is not always desirable because it can sometimes be disastrous in mental disorders such as epilepsy.[7] Electrical engineers apply desynchronization to implement time division multiple access (TDMA) that prevents message collisions and provides asynchronous sleep cycles for nodes on wireless sensor networks.[8] Indeed, considering the enormous number of connections between cells and organs in the body, the apparent independence or desynchronization between rhythms in different organs may be a bigger puzzle rather than their synchronization.[9] Complex systems sometimes show partial synchronization. Two types of partial synchronization have been recognized. The *chimera state*

a) These authors are equally contributed.

b) Author to whom correspondence. Electronic mail: jojunghyo@apctp.org

shows spatial separation of synchronized and desynchronized domains,[10] whereas the *periodic synchronization* shows temporal alternation between synchronized and desynchronized states.[11, 12]

Therefore, context-dependent (de)synchronization and their controllability are necessary to ensure complex systems function effectively. In particular, to disrupt undesired synchronization, various methods have been proposed. These include simply decreasing interaction strengths to stimulating with a short pulse,[13] and giving linear/nonlinear delayed feedback.[14, 15] However, these schemes have been applied for desynchronization in globally coupled networks. Biological networks are generally more complex. They have (i) hierarchical organizations with local/global coupling,[16-18] (ii) heterogeneous coupling signs with excitatory/inhibitory connections,[19] and (iii) dynamic coupling amplitudes depending on the activities of network elements.[20] Thus desynchronization may not be a simple task in such complex networks.

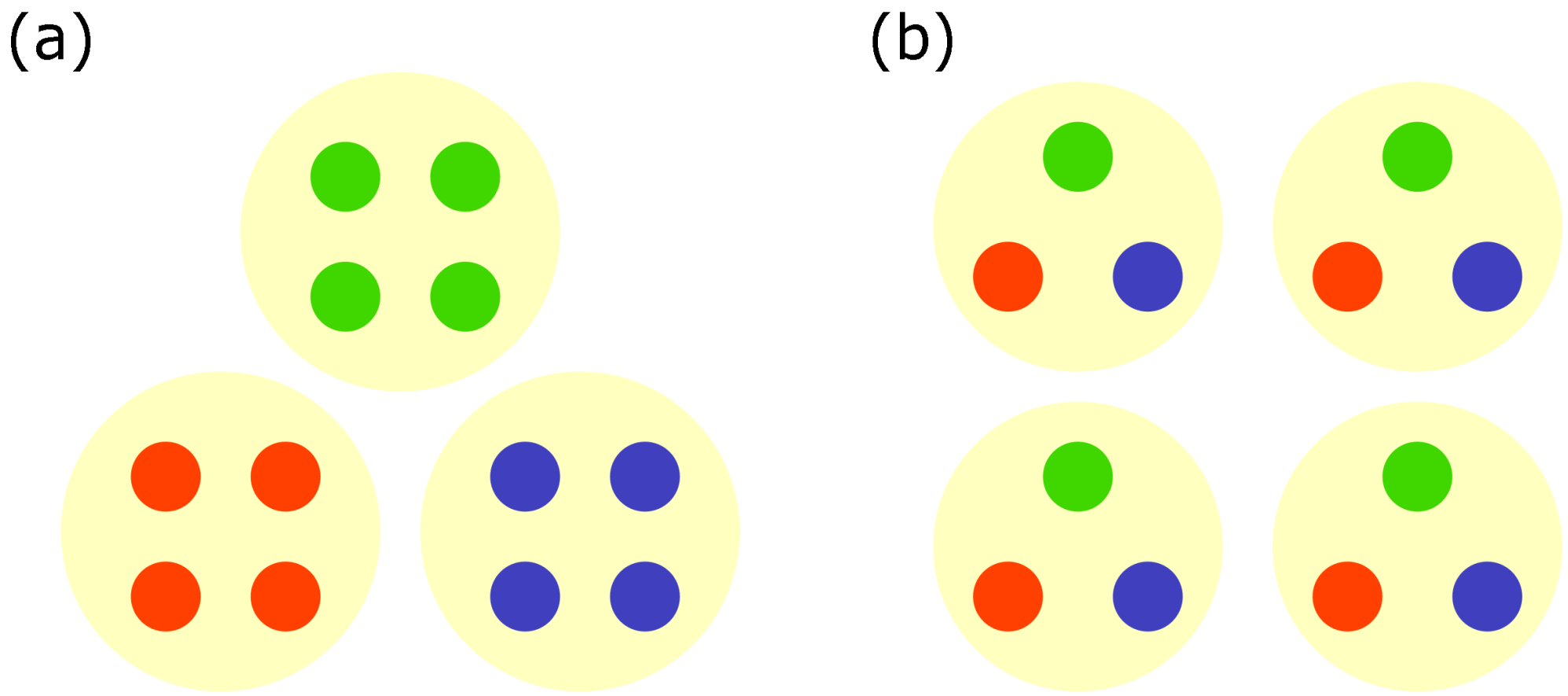

FIG. 1. (Color online) Two hierarchical organizations of twelve oscillators. (a) Three heterogeneous subsystems (yellow circles) have distinct oscillators (green, red, and blue circles), but each subsystem has homogeneous oscillators. (b) Four homogeneous subsystems (yellow circles) have the same composition of three heterogeneous oscillators (green, red, and blue circles).

Recent studies have investigated the control of (de)synchronization in locally connected networks: introducing more inhibitory couplings can suppress synchronization in the complex networks;[21] tuning coupling phases (or delay) and/or balancing between excitatory and inhibitory nodes can switch between synchronization and desynchronization;[22, 23] optimizing coupling signs and amplitudes can possibly induce cluster synchronization, in which groups of oscillators are separately synchronized;[24] and changing mesoscale network motifs can also lead to different synchronization patterns.[25]

In this study, we consider hierarchically-organized oscillators and their synchronization. A hierarchical organization of heterogeneous subsystems with homogeneous oscillators (Fig. 1(a)) has been analytically studied.[26, 27] Here, we focus on another hierarchical organization of homogeneous subsystems with heterogeneous oscillators (Fig. 1(b)), because many biological and social systems have the latter hierarchy. The pancreas is composed by islets of Langerhans in which heterogeneous endocrine cells interact with each other;[28] and a country is composed by cities in which people with different opinions interact with each other.[12] With this hierarchy in mind, we propose an effective strategy to control the synchronization between subsystems, based on a simple hypothesis that hierarchical systems become increasingly difficult to generate coherent behavior as the complexity of subsystems increases. We examine how the network motif (coupling signs between oscillators) of a subsystem affects its local complexity and the global synchronization between subsystems. To probe this hypothesis, we formulate a minimal model using coupled Stuart-Landau oscillators, and demonstrate that the subsystem complexity can predict the global synchronization of the hierarchical system.

## II. MODEL

### A. Subsystem dynamics and local complexity

To formulate synchronization phenomena of hierarchically-networked complex oscillators (FIG. 1(b)), we adopt the Stuart-Landau oscillators in which amplitudes and phases are variables:[29]

$$\dot{Z}_\sigma = \left(\lambda_\sigma - |Z_\sigma|^2 + i\omega_\sigma\right) Z_\sigma + K \sum_{\sigma' \neq \sigma} a_{\sigma\sigma'} Z_{\sigma'} . \tag{1}$$

First, we consider a *subsystem* that consists of three oscillators with $\sigma, \sigma' \in \{1,2,3\}$ as a minimal set for generating different degrees of complexity. The complex variable $Z_\sigma \equiv r_\sigma e^{i\theta_\sigma}$ combines the amplitude $r_\sigma$ and the phase $\theta_\sigma$ for $\sigma$ oscillator, and the parameters $\lambda_\sigma$ and $\omega_\sigma$ control its stationary amplitude and intrinsic frequency, respectively. Then, $K$ and $a_{\sigma\sigma'}$ represent the coupling strength and sign from $\sigma'$ oscillator to $\sigma$ oscillator. If one considers a simpler system of which subsystems consist of only two oscillators, the system generates simple in-phase or out-of-phase relations between the two oscillators except for the special condition that they have asymmetric couplings with equal strengths (Fig. S1). The dynamics of the complex variables $Z_\sigma \equiv r_\sigma e^{i\theta_\sigma}$ can be understood by decomposing amplitude $r_\sigma$ and $\theta_\sigma$ phase parts:

$$\dot{r}_\sigma = \left(\lambda_\sigma - r_\sigma^2\right) r_\sigma + K \sum_{\sigma' \neq \sigma} a_{\sigma\sigma'} r_{\sigma'} \cos\left(\theta_{\sigma'} - \theta_\sigma\right), \tag{2}$$

$$\dot{\theta}_\sigma = \omega_\sigma + K \sum_{\sigma' \neq \sigma} a_{\sigma\sigma'} \frac{r_{\sigma'}}{r_\sigma} \sin\left(\theta_{\sigma'} - \theta_\sigma\right). \tag{3}$$

In the absence of coupling ($K = 0$), the amplitude converges to a stable focus ($r_\sigma = 0$) for $\lambda_\sigma < 0$, but the focus loses stability at the Hopf bifurcation point ($\lambda_\sigma = 0$), and a stable limit-cycle emerges with amplitude $\sqrt{\lambda_\sigma}$ and frequency $\omega_\sigma$ for $\lambda_\sigma > 0$. Note that $r_\sigma = -\sqrt{\lambda_\sigma}$ is another solution for $\lambda_\sigma > 0$. Here we use only positive-definite $r_\sigma$; when $r_\sigma$ becomes negative in simulations, we make it positive with the transformation, $r_\sigma \to -r_\sigma$ and $\theta_\sigma \to \theta_\sigma + \pi$.

Once the coupling is applied ($K > 0$), the subsystem produces rich dynamics depending on the adjacency matrix $\boldsymbol{A}$ with entries $a_{\sigma\sigma'}$ that determine the coupling sign from $\sigma'$ oscillator to $\sigma$ oscillator. The phase part of Eq. (3) is a generalized Kuramoto model,[29] in which oscillators have positive and negative couplings, and their coupling strengths depend on their amplitudes. The amplitude dependence has the physical meaning that when oscillator $\sigma'$ affects $\sigma$ oscillator, the coupling strength is proportional to the amplitude of affector $r_{\sigma'}$, but inversely proportional to the amplitude of receiver $r_\sigma$; i.e., the pair of strong affector and weak receiver exhibits a maximal coupling.

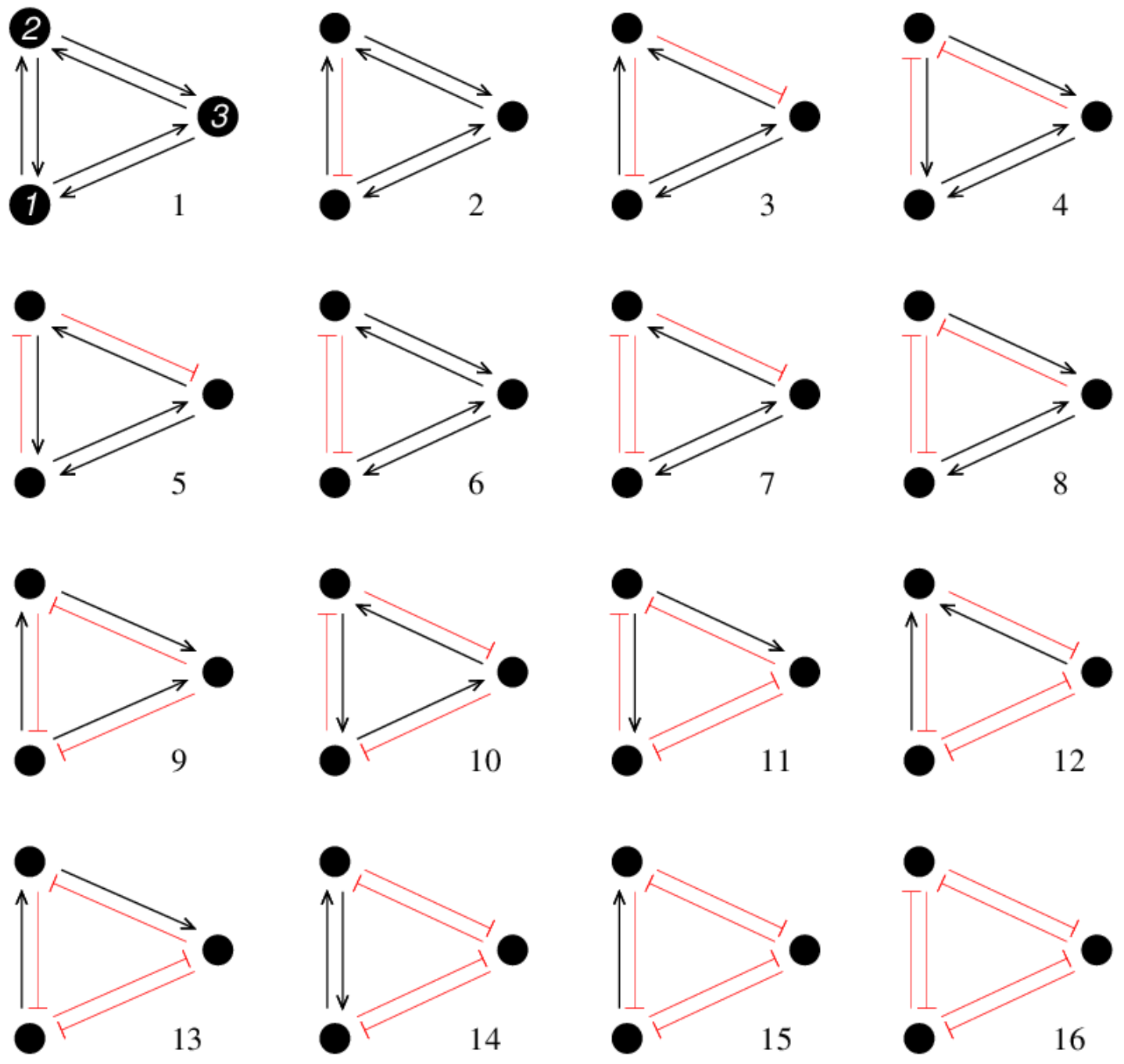

FIG. 2. (Color online) All possible coupling signs (network motifs) given three oscillators within a subsystem. Black arrows: positive interactions; red bar-headed arrows: negative interactions.

When the coupling is weak, the amplitudes can be approximated to $r_\sigma \approx \sqrt{\lambda_\sigma}$, and the three phase equations of Eq. (3) can be reduced to two equations of the phase differences, $x \equiv \theta_1 - \theta_2$ and $y \equiv \theta_1 - \theta_3$. Assuming identical intrinsic frequencies ($\omega_\sigma = \omega$) for simplicity, we obtain

$$\dot{x} = -(b_{12} + b_{21})\sin x - b_{13}\sin y - b_{23}\sin(x - y), \quad (4)$$

$$\dot{y} = -b_{12}\sin x - (b_{13} + b_{31})\sin y + b_{32}\sin(x - y), \quad (5)$$

where $b_{\sigma\sigma'} \equiv a_{\sigma\sigma'}Kr_{\sigma'}/r_{\sigma}$. Their steady states ($\dot{x}=0$ and $\dot{y}=0$) are governed by the parameters $\lambda_\sigma$ and the adjacency matrix $\boldsymbol{A}$. Here self-couplings can be safely ignored ($a_{\sigma\sigma}=0$), because their contribution is absent in Eq. (3). Considering that the off-diagonal elements $a_{\sigma\sigma'}$ for $\sigma \neq \sigma'$ can take either 1 or -1 for positive or negative couplings, the adjacency matrix $\boldsymbol{A}$ has a total of 64 (= $2^6$) possibilities. Leaving index degeneracy aside, 16 cases remain (FIG. 2). Most of them drive ($x$, $y$) to single or double fixed points at steady states, regardless of $\lambda_\sigma$ (Fig. S2). However, two anti-symmetric matrices Network 9 and Network 10,

$$A = \begin{pmatrix} 0 & -1 & -1 \\ 1 & 0 & -1 \\ 1 & 1 & 0 \end{pmatrix}, \begin{pmatrix} 0 & 1 & -1 \\ -1 & 0 & 1 \\ 1 & -1 & 0 \end{pmatrix}$$

are exceptional in that they produce multiple fixed points for similar $\lambda_\sigma(\lambda_1 \approx \lambda_2 \approx \lambda_3)$. Moreover, the two anti-symmetric matrices also generate single fixed points for largely dissimilar $\lambda_\sigma$. Note that the effective coupling $b_{\sigma\sigma'} \equiv a_{\sigma\sigma'}Kr_{\sigma'}/r_{\sigma}$ can be different from the topological coupling $a_{\sigma\sigma'}$ for largely dissimilar $\lambda_\sigma$. Therefore, we focus more on these two subsystems that can alter their *complexity* by changing $\lambda_\sigma$. In this study, we define the complexity of subsystems based on the number of stable fixed points and demonstrate that the complexity can predict the synchronization of the hierarchically-networked complex oscillators.

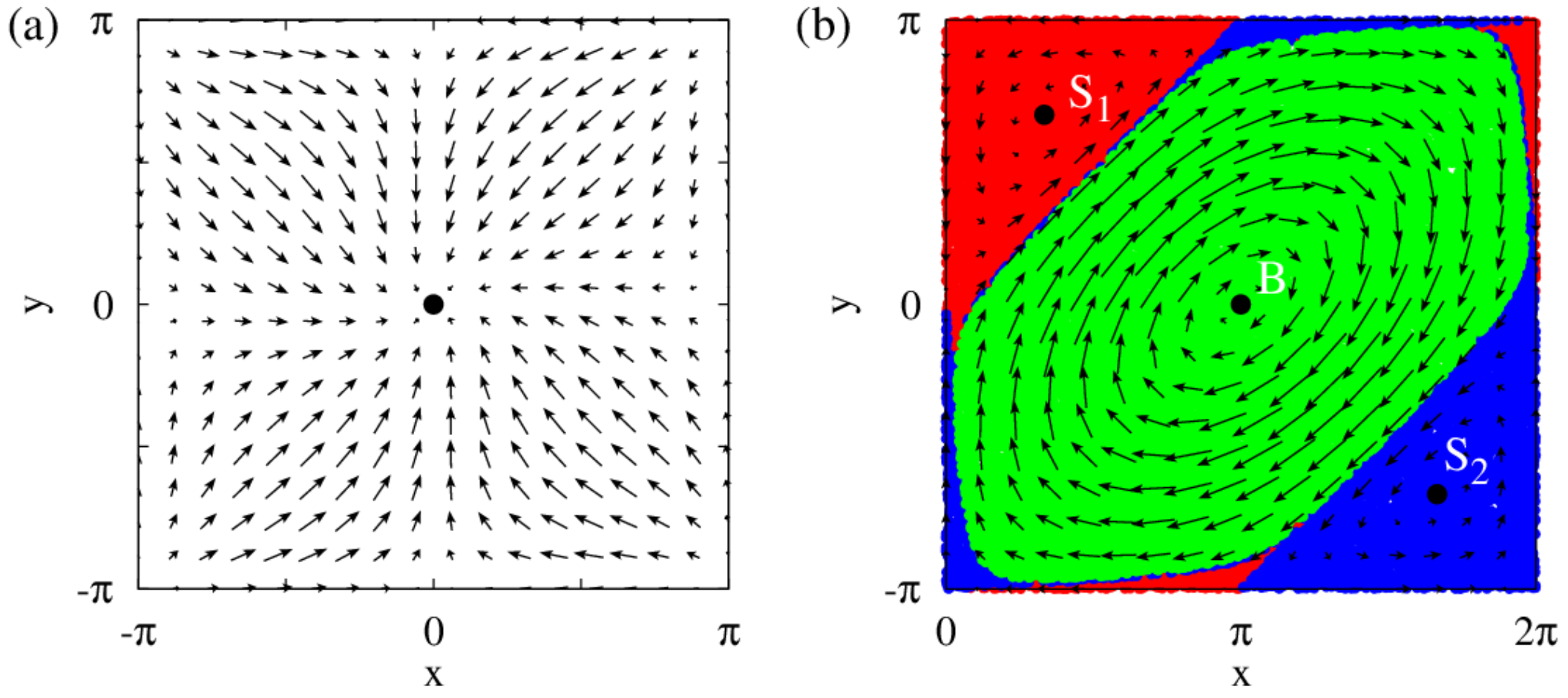

FIG. 3. (Color online) Attractors of three coupled oscillators. Stable fixed points (black points) and vector flows ($\dot{x}, \dot{y}$) on the plane ($x, y$), where $x \equiv \theta_1 - \theta_2$ and $y \equiv \theta_1 - \theta_3$ are relative phases between three oscillators. Depending on the parameter values of $\lambda_\sigma$ for determining the activities of oscillators, anti-symmetrically coupled three oscillators produce a single attractor for (a) $\lambda_1 = 1.0$ and $\lambda_2 = \lambda_3 = 0.2$ or three attractors for (b) $\lambda_1 = \lambda_2 = \lambda_3 = 1.0$. The basins of the three attractors, $S_1$, $S_2$, and $B$, are painted with red (top left), blue (bottom right), and green (middle), respectively. For these plots, we used a coupling strength ($K = 0.1$), identical intrinsic frequency ($\omega_\sigma = 1$) and signs ($a_{11} = a_{22} = a_{33} = 0$, $a_{12} = a_{13} = a_{23} = -1$, $a_{21} = a_{31} = a_{32} = 1$), where $a_{\sigma\sigma'}$ represents the interaction sign from $\sigma'$ oscillator to $\sigma$ oscillator.

The first anti-symmetric subsystem (Network 9) has three distinguishable oscillators: The first oscillator attracts the other two; the second repels the other two; and the third attracts one and repels one. In contrast, the second anti-symmetric subsystem (Network 10) has three indistinguishable oscillators. Network 9 produces single fixed points for dissimilar $\lambda_\sigma$ (FIG. 3(a)), but three fixed points for similar $\lambda_\sigma$ (FIG. 3(b)). For $\lambda_1 = \lambda_2 = \lambda_3 = 1.0$, the $(x, y)$ phase plane has three fixed points: (π/3, 2π/3), (5π/3, -2π/3), and (π, 0). We label them $S_1$, $S_2$, and $B$, because $S_1$ and $S_2$ have small basins of attractors, whereas $B$ has a big basin. When the amplitude dynamics in Eq. (2) is off ($\dot{r}_\sigma = 0$), and the phase model in Eq. (3) is only considered with frozen amplitudes, the subsystem is not attracted to the centers ($S_1$, $S_2$, and $B$), but revolves endlessly around the centers at given initial radii.[28] $S_1$, $S_2$, and $B$ are special positions in which the coupling terms, $\sum_{\sigma' \neq \sigma} a_{\sigma\sigma'} r_{\sigma'} \cos(\theta_{\sigma'} - \theta_\sigma) = 0$ in Eq. (2), vanish and amplitudes $r_\sigma = \sqrt{\lambda_\sigma}$ become fixed. Network 10 has the same phase plane as Network 9 with simple translations $x \to x - \pi$ and $y \to y$.

Simple counting of the number of fixed points can represent the local complexity of subsystems depending on $\lambda_\sigma$ (Fig. S2). However, if different fixed points are realized with different frequencies, the simple counting is not enough to consider the weighted realization. As an extreme example, suppose a subsystem has multiple fixed points where one has a large basin of attraction, but the others has tiny basins. Then, the subsystem should have low complexity despite of the multiple fixed points. Therefore, to consider the realization uncertainty of different fixed points, we quantified the local complexity by using Shannon entropy:

$$S = -\sum_{\mu} P_\mu \log_2 P_\mu \tag{6}$$

where $\mu$ represents different fixed points, and $P_\mu$ is their realization frequency. This measures the local complexity of subsystems depending on $\lambda_\sigma$ (FIG. 4). Note that the complexity map is identical for the two anti-symmetric subsystems (Networks 9 and 10).

### B. Coupling subsystems and global synchronization

To demonstrate that the global synchronization between subsystems can depend on the local complexity of subsystems, we construct a hierarchical system composed of multiple subsystems. Each subsystem has three coupled oscillators. The amplitude and phase of the oscillator $\sigma$ in the $n$th subsystem are represented as $Z_{n\sigma} \equiv r_{n\sigma} e^{i\theta_{n\sigma}}$. We consider an *inter-subsystem* coupling between subsystems in addition to the *intra-subsystem* coupling between oscillators within a subsystem (FIG. 1(b)). In particular, we start by simply copying the intra-subsystem coupling into the inter-subsystem coupling:

$$\dot{Z}_{n\sigma} = \left(\lambda_\sigma - |Z_{n\sigma}|^2 + i\omega_{n\sigma}\right) Z_{n\sigma} + K \sum_{\sigma' \neq \sigma} a_{\sigma\sigma'} Z_{n\sigma'} + \varepsilon K \sum_{m \neq n} \sum_{\sigma' \neq \sigma} a_{\sigma\sigma'} Z_{m\sigma'}, \tag{7}$$

where $\varepsilon$ is introduced to represent weaker strengths of inter-subsystem couplings compared with intra-subsystem couplings ( $0 < \varepsilon < 1$ ). By using an arithmetic average $Z_{n\sigma}$ of for $N$ subsystems,

$$\bar{Z}_\sigma \equiv \frac{1}{N} \sum_{n=1}^{N} Z_{n\sigma}, \tag{8}$$

we rearrange Eq. (7) as

$$\dot{Z}_{n\sigma} = \left(\lambda_\sigma - |Z_{n\sigma}|^2 + i\omega_{n\sigma}\right) Z_{n\sigma} + \tilde{K} \sum_{\sigma' \neq \sigma} a_{\sigma\sigma'} \left(Z_{n\sigma'} + \tilde{\varepsilon}\, \bar{Z}_{\sigma'}\right), \tag{9}$$

where $\tilde{K} = K(1-\varepsilon)$ and $\tilde{\varepsilon} = N\varepsilon/(1-\varepsilon)$. Henceforth, we use the rescaled parameters as $\tilde{K} \to K$ and $\tilde{\varepsilon} \to \varepsilon$ . This equation can be interpreted as the mean fields $\bar{Z}_\sigma$ affect every subsystem. To probe the global synchronization between subsystems, we define order parameters for three oscillators:

$$\rho_\sigma = \left| \frac{1}{N} \sum_{n=1}^{N} e^{i\theta_{n\sigma}} \right| \tag{10}$$

with $\sigma \in \{1,2,3\}$. Here we use the average of the three oscillators to present the degree of synchronization, and we represent them simply as $\rho = (\rho_1+\rho_2+\rho_3)/3$, unless otherwise specified. For the anti-symmetric intra-subsystem couplings, the arithmetic mean field leads to desynchronize subsystems ( $\rho = 0$ ), because the inter-subsystem coupling in Eq. (9) effectively generates repulsive interactions between subsystems. For an example of two-subsystems systems ($N$ = 2), $Z_{1\sigma}$ indirectly interacts with $Z_{2\sigma}$ through $Z_{2\sigma} \to Z_{2\sigma'} \to \bar{Z}_{\sigma'} \to Z_{1\sigma}$ or $Z_{1\sigma} \to Z_{1\sigma'} \to \bar{Z}_{\sigma'} \to Z_{2\sigma}$. These three-step interactions always yield a negative loop as a net for the anti-symmetric matrices $\boldsymbol{A}$. The effective repulsion between subsystems leads them to stay as far away as possible. This state has been referred as the *splay state*.[24] Therefore, under the arithmetic mean field, the anti-symmetric intra-subsystem coupling can provide an effective scheme for the desynchronization of hierarchical systems (Fig. S3).

For the inter-subsystem coupling, we also consider a geometric average (log-average) of $Z_{n\sigma}$,

$$\bar{Z}_\sigma \equiv \bar{r}_\sigma e^{i\bar{\theta}_\sigma} = \left[ \prod_{n=1}^{N} Z_{n\sigma} \right]^{\frac{1}{N}}. \tag{11}$$

The geometric average that counts large outliers less important than the arithmetic average is frequently suitable in biological systems.[30] Unlike the arithmetic average in Eq. (8), the geometric average decouples amplitude and phase averages:

$$\bar{r}_\sigma \equiv \left[\prod_{n=1}^{N} r_{n\sigma}\right]^{\frac{1}{N}}, \tag{12}$$

$$\bar{\theta}_\sigma \equiv \frac{1}{N}\sum_{n=1}^{N}\theta_{n\sigma}. \tag{13}$$

The average phase $\bar{\theta}_\sigma$ is given as an arithmetic average of bare phases, independent on amplitudes. After adopting the geometric mean field $\bar{Z}_\sigma$ of Eq. (11), we decompose Eq. (9) into amplitude and phase parts:

$$\dot{r}_{n\sigma} = (\lambda_\sigma - r_{n\sigma}^2)\, r_{n\sigma} + K\sum_{\sigma'\neq\sigma} a_{\sigma\sigma'}\left[ r_{n\sigma'}\cos(\theta_{n\sigma'}-\theta_{n\sigma}) + \varepsilon\,\bar{r}_{\sigma'}\cos(\bar{\theta}_{\sigma'}-\theta_{n\sigma})\right], \tag{14}$$

$$\dot{\theta}_{n\sigma} = \omega_{n\sigma} + K\sum_{\sigma'\neq\sigma} a_{\sigma\sigma'}\left[ \frac{r_{n\sigma'}}{r_{n\sigma}}\sin(\theta_{n\sigma'}-\theta_{n\sigma}) + \varepsilon\frac{\bar{r}_{\sigma'}}{r_{n\sigma}}\sin(\bar{\theta}_{\sigma'}-\theta_{n\sigma})\right]. \tag{15}$$

In this study, we adopt the geometric mean-field coupling, because it produces broader spectrum of synchronization between subsystems. This completes our formulation of hierarchical oscillators.

## III. RESULTS

### A. Local complexity and global synchronization

We now investigate the global synchronization of the hierarchically-organized oscillators governed by Eqs. (14) and (15). We consider a sufficiently large system ($N$ = 100) of which dynamics is insensitive to initial conditions. In the absence of the inter-subsystem coupling ($\varepsilon = 0$), the two anti-symmetric subsystems (Networks 9 and 10) have single fixed points of $\theta_{n1}-\theta_{n2}$ and $\theta_{n1}-\theta_{n3}$ for largely dissimilar $\lambda_\sigma$ (Fig. 3(a)). Therefore, individual subsystems should be attracted into the identical fixed point. Then they have low complexity index ($S$ = 0) by Eq. (6). Once the inter-subsystem coupling is turned on ($\varepsilon = 0.1$), the subsystems become easily synchronized ($\rho \approx 1$) under the low subsystem complexity. In contrast, the global synchronization is hindered under a high subsystem complexity. When the subsystem has multiple fixed points $S \approx 2$ for similar $\lambda_\sigma$, individual subsystems usually enter different basins of attractors (Fig. 3(b)). Then they become difficult to be synchronized, or easy to be desynchronized. This observation concludes that the subsystem complexity $S$ can predict the system synchronization $\rho$. Figure 4 clearly demonstrates the correlation between the subsystem complexity and the global synchronization in a large parameter space of $\lambda_\sigma$ for the 16 network motifs of subsystems. Note that measuring the subsystem complexity $S$ does not require a heavy computation for simulating the whole system. Therefore, the local complexity ($S$) can be a useful index for predicting the global synchronization ($\rho$). The Pearson correlation coefficients between $S$ and $\rho$ are -0.81 (Network 9), -0.85 (Network 10), and -0.71 (all 16 networks) for a weak inter-subsystem coupling ( $\varepsilon = 0.1$ ) (Fig. S4(a)).

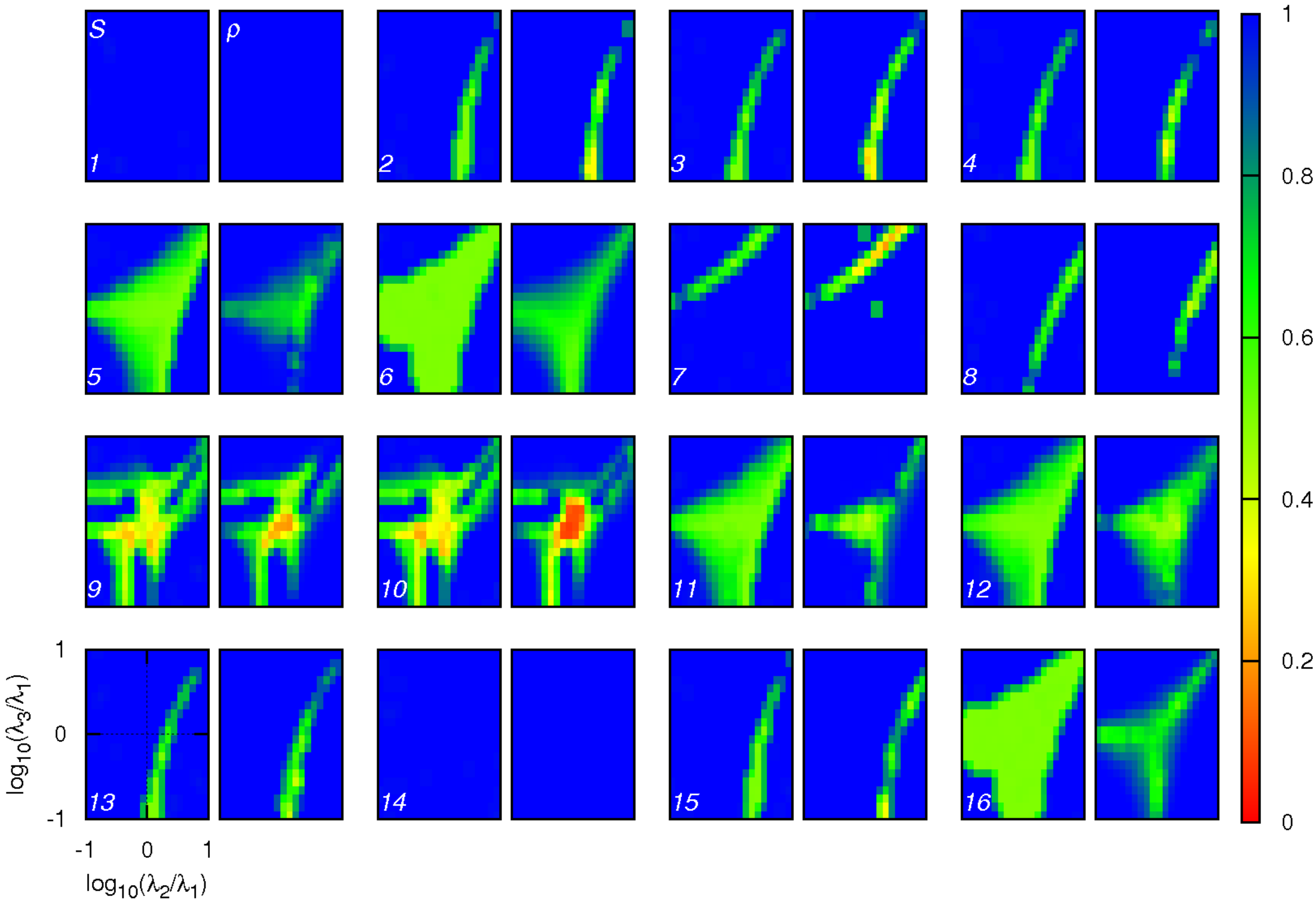

FIG. 4. (Color online) Local complexity and global synchronization. The local complexity of subsystems ($S$, left panels) is measured by the Shannon entropy of realized fixed points ($x^*$, $y^*$) of three coupled oscillators within a certain accuracy ( $\delta x^*, \delta y^* < \pi/5$ ), where $x \equiv \theta_1 - \theta_2$ and $y \equiv \theta_1 - \theta_3$ are relative phases between three oscillators. One hundred ensembles are used to obtain the frequencies of realized fixed points. The average degree of synchronization ($\rho$, right panels) between subsystems is measured. For clear comparison with $\rho$ (red: desynchronization; blue: synchronization), the complexity index $S$ is reduced by half (red: 1; blue: 0). Axes have base-10-logarithmic scales to allow presentation of a large range of $\lambda_\sigma$ , and each axis has 21 bins. For these plots, we used an intra-subsystem coupling ( $K = 0.1$ ), an inter-subsystem coupling ( $\varepsilon = 0.1$ ), and identical intrinsic frequency ( $\omega_{n\sigma} = 1$ ) for $N$ = 100 subsystems.

## B. Local coupling signs between three oscillators and their phase relations

We further examine the phase relations between three oscillators of the 16 networks. The phase differences between $\theta_{n1}$, $\theta_{n2}$ and $\theta_{n3}$ can be summarized by a single index, $P = \langle 2\cos(\theta_{n1} - \theta_{n2}) + \cos(\theta_{n1} - \theta_{n3}) \rangle$, which is an ensemble average over the subsystem index $n$. In particular, we focus on four distinct states: (i) $P$ = 3 ( $\theta_{n1} = \theta_{n2} = \theta_{n3}$ ); (ii) $P$ = 1 ( $\theta_{n1} = \theta_{n2} = \theta_{n3} + \pi$ ); (iii) $P$ = -1 ( $\theta_{n1} = \theta_{n2} + \pi = \theta_{n3}$ ); (iv) $P$ = -3 ( $\theta_{n1} + \pi = \theta_{n2} = \theta_{n3}$ ) considering the $2\pi$ periodicity. The 16 networks showed different patterns of the phase states depending on the parameter values of $\lambda_\sigma$ :

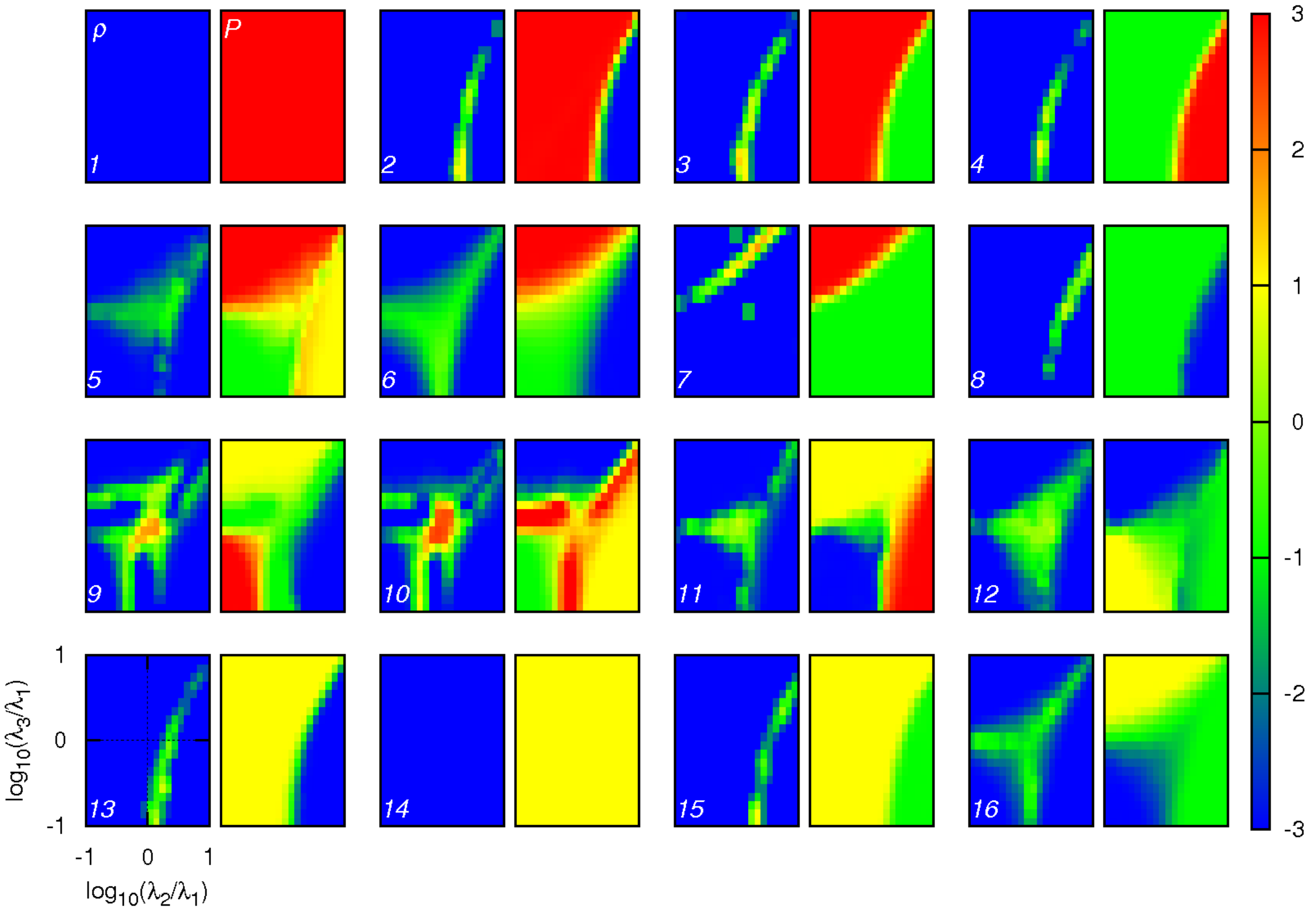

FIG. 5. (Color online) Phase relations between three oscillators for 16 networks. Right panels: index $P = \langle 2\cos(\theta_{n1} - \theta_{n2}) + \cos(\theta_{n1} - \theta_{n3}) \rangle$ for representing the phase relations between three oscillators, where $\langle \ldots \rangle$ indicates an average over the subsystem index $n \in \{1,\ldots,N\}$. Four distinct states exist: (i) $P$ = 3 (red) for $\theta_{n1} = \theta_{n2} = \theta_{n3}$; (ii) $P$ = 1 (yellow) for $\theta_{n1} = \theta_{n2} = \theta_{n3} + \pi$; (iii) $P$ = -1 (green) for $\theta_{n1} = \theta_{n2} + \pi = \theta_{n3}$; (iv) $P$ = -3 (blue) for $\theta_{n1} + \pi = \theta_{n2} = \theta_{n3}$. Left panels: index $\rho$ for the global synchronization between subsystems, the same plot in FIG. 4. Note that the scale of $\rho$ is between 0 (desynchronization, red) and 1 (synchronization, blue), and its scale bar is not shown here. Axes have base-10-logarithmic scales to allow presentation of a large range of $\lambda_\sigma$, and each axis has 21 bins. For these plots, we used an intra-subsystem coupling ($K = 0.1$), an inter-subsystem coupling ($\varepsilon = 0.1$), and identical intrinsic frequency ($\omega_{n\sigma} = 1$) for $N$ = 100 subsystems.

**1. One phase state.** Networks 1 and 14 showed full synchronization over the entire parameter space (FIG. 4). However, they have different phase patterns, although they are homogeneous over the parameter space (Fig. 5). Network 1 had in-phase relations between all three oscillators ($P$ = 3), while Network 14 had an in-phase relation between $\theta_{n1}$ and $\theta_{n2}$, but an out-of-phase relation between $\theta_{n1}$ and $\theta_{n3}$ ($P$ = 1) (Fig. S5).

**2. Two phase states.** Networks 2, 3, 4, 7, 8, 13, and 15 showed two phase states separated by a thin transition region (Fig. 5 and Fig. S6). The coexistence of multiple phase states in the transition region hindered the global synchronization between subsystems.

**3. Three phase states.** Networks 5, 6, 11, 12, and 16 showed three phase states (Fig. 5). At $\lambda_1 = \lambda_2 = \lambda_3$, they had different synchronization patterns. Networks 5, 11, and 12 showed cluster synchronization states, in which oscillators 1, 2, and 3 in different subsystems were grouped into two synchronized clusters, respectively (Fig. S7(a-c)). On the other hand, Network 6 had two synchronized clusters for oscillators 1 and 2, but a single synchronized cluster for oscillator 3 (Fig. S7(d)). The states of cluster synchronization correspond to the traveling wave state.[31] However, Network 16 did not show clear cluster synchronization (Fig. S7(e)).

**4. Four phase states.** Networks 9 and 10, constructed by anti-symmetric couplings between any pairs of oscillators, showed unique synchronization patterns. They generated four distinct phase states depending on the parameter values of $\lambda_\sigma$. In particular, they had different phase states in three parameter regions (Fig. 5): (i) $\lambda_1$-dominant region (left bottom in panels); (ii) $\lambda_2$-dominant region (right bottom, $\lambda_2 >> \lambda_1 >> \lambda_3$); and (iii) $\lambda_3$-dominant region (left top, $\lambda_3 >> \lambda_1 >> \lambda_2$). The three distinct states met at $\lambda_1 = \lambda_2 = \lambda_3$, in which the coexistence of multiple states showed a maximal desynchronization of the hierarchical system.

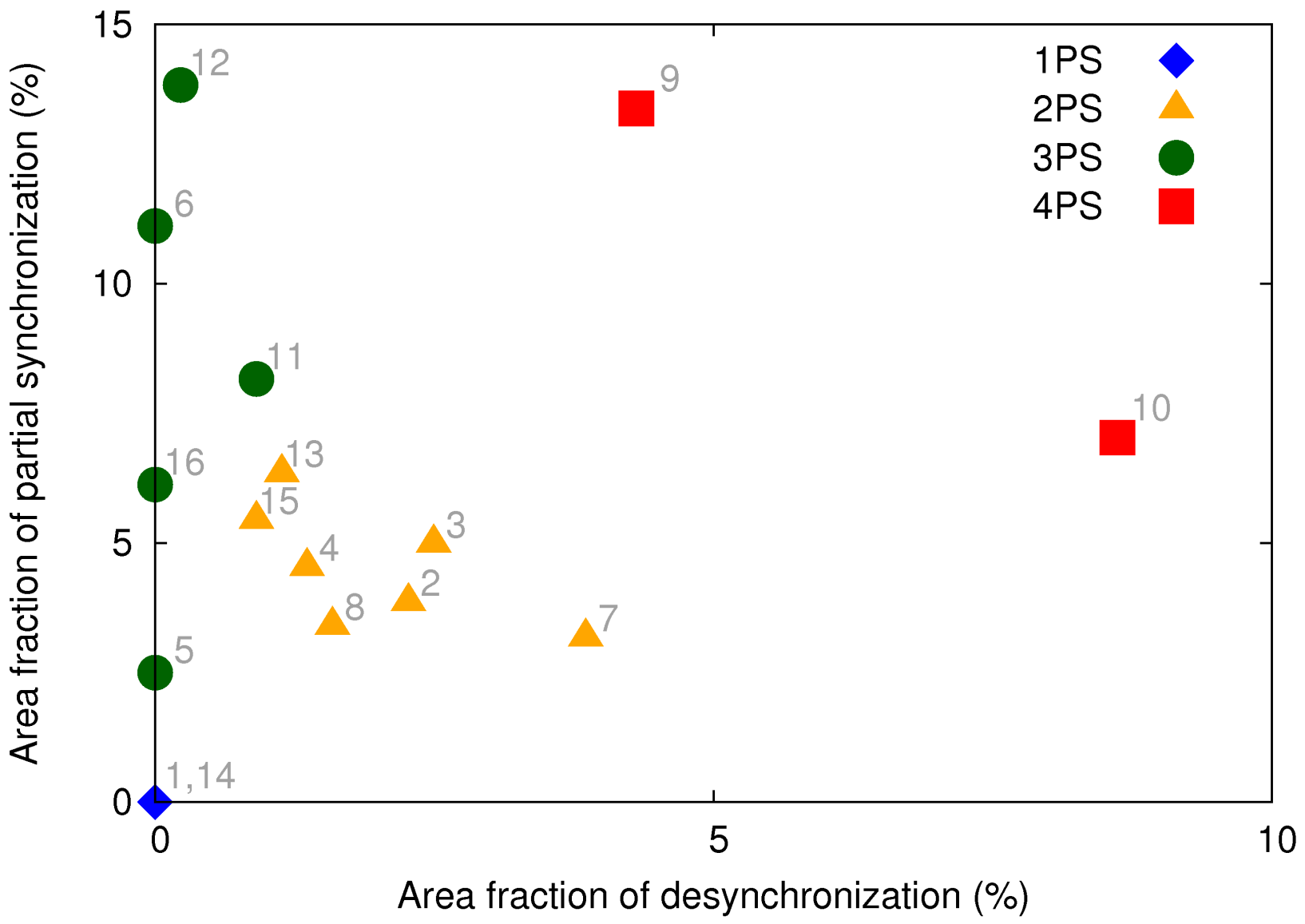

FIG. 6. (Color online) Synchronizability of 16 networks. The degree of synchronization ($\rho = 1$, perfect synchronization; $\rho = 0$, perfect desynchronization) between subsystems depends on activity parameters $\lambda_\sigma$ for three oscillators $\sigma \in \{1,2,3\}$. The coupling signs between three oscillators define 16 networks. The whole parameter space is grouped into three areas for $\rho < 1/3$ (desynchronization), $1/3 < \rho < 2/3$ (partial synchronization), and $\rho > 2/3$ (synchronization). The area fractions for desynchornization versus partial synchronization are plotted. The 16 networks have different numbers of phase states (PS), which are defined by the phase relations between three oscillators.

Finally, we quantified the synchronizability of the 16 networks by measuring the parameter regions in Fig. 4 for desynchronization ($\rho < 1/3$) and partial synchronization ($1/3 < \rho < 2/3$). The area fractions could categorize the 16 networks (Fig. 6), of which grouping was consistent with the above categorization of phase states. Networks 9 and 10 of four phase states had large regions for desynchronization and partial synchronization, while Networks 1 and 14 of one phase state had no region for them. The networks of two phase states had relatively small regions for them, and the network of three phase states showed wide ranges of partial synchronization, but negligible region for desynchronization.

### C. Limited predictability of global synchronization by local complexity

The strong correlation between local complexity and global synchronization did not always hold. Strong inter-subsystem couplings, large heterogeneities in intrinsic frequencies, or noisy fluctuations of phases decreased the correlation.

**1. Strong inter-subsystem coupling.** If the inter-subsystem coupling was absent ( $\varepsilon = 0$ ), then subsystems always remained desynchronized as they do not interact with each other. As the inter-subsystem coupling increased ( $\varepsilon = 0.1$ ), subsystems started to synchronize, and the global synchronization patterns had a strong correlation with the local complexity of subsystems. However, when we further increased the inter-subsystem coupling strength ( $\varepsilon > 1$ ), subsystems became fully synchronized regardless of the network motifs of subsystems (Fig. S8). We examined how the inter-subsystem coupling strength affected the correlation between local complexity and global synchronization (Fig. 7). We found that the correlation was maximal at a weak inter-subsystem coupling ( $\varepsilon = 0.1$ ) for oscillators with identical intrinsic frequencies in the absence of noise.

**2. Heterogeneous intrinsic frequencies.** As the intrinsic frequencies $\omega_{n\sigma}$ of oscillators were more heterogeneous, the global synchronization between subsystems was expected to decrease. To investigate this limitation, we distributed the intrinsic frequencies to follow a uniform random distribution, $\omega_{n\sigma} \in [\omega_0 - \Delta\omega, \omega_0 + \Delta\omega]$, around the central frequency $\omega_0 = 1$ . Indeed largely heterogeneous intrinsic frequencies ( $\Delta\omega > 0.1$ ) prevented the global synchronization between subsystems regardless of their network motifs given a weak inter-subsystem coupling ( $\varepsilon = 0.1$ ). Gaussian random distributions of $\omega_{n\sigma}$ showed similar results. However, the local complexity was still correlated with the global synchronization when the heterogeneity was small ( $\Delta\omega = 0.01$ ), although the correlation was reduced (Fig. 7). Then, their maximum correlation appeared at a higher inter-subsystem coupling strength ( $\varepsilon \approx 0.2$ ) than the optimal coupling strength ( $\varepsilon = 0.1$ ) for identical oscillators ( $\Delta\omega = 0$ ).

**3. Noise.** To examine the noisy phase fluctuations of oscillators, we added a white noise $\xi(t)$ to the phase dynamics in Eq. (15). The white noise satisfies zero mean $\langle \xi(t) \rangle = 0$ and delta correlated variance $\langle \xi(t)\xi(t') \rangle = D\delta(t-t')$, where $D$ is the noise intensity and $\delta(t-t')$ is the Dirac delta function that is zero everywhere except at $t = t'$. As the noise intensity increased, the system became more desynchronized in general. Too strong noise (e.g., $D$ = 0.09) made any subsystems desynchronized except for Network 1 ( $\rho = 0.36$ on average). However, under a weak noise ($D$ = 0.0025), the correlation between local complexity and global synchronization still existed, although it was reduced. (Fig. 7).

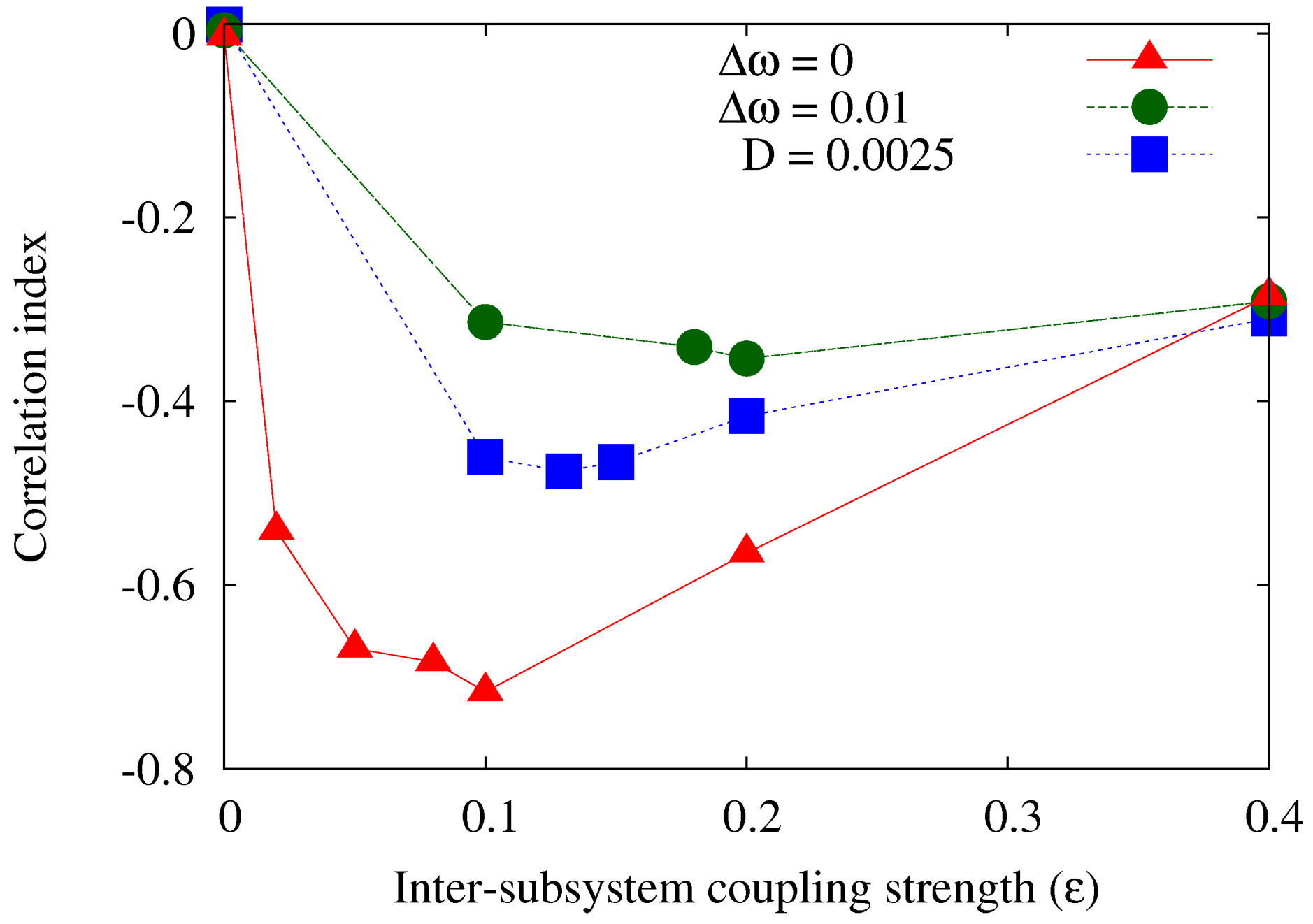

FIG. 7. Pearson correlations between local complexity and global synchronization for various inter-subsystem coupling strengths with heterogeneity and noise. Oscillators in each subsystem have homogeneous ( $\omega_{n\sigma} = 1$ with $\Delta\omega = 0$ ) or heterogeneous intrinsic frequencies ( $\omega_{n\sigma} \in [0.99, 1.01]$ with $\Delta\omega = 0.01$ ), and their phase dynamics has a white noise $\xi(t)$ satisfying $\langle \xi(t) \rangle = 0$ and $\langle \xi(t), \xi(t') \rangle = D\delta(t-t')$. Plotted are three cases: (i) $\Delta\omega = 0$ and $D$ = 0 (red triangles); (ii) $\Delta\omega = 0.01$ and $D$ = 0 (green circles); and (iii) $\Delta\omega = 0$ and $D$ = 0.0025 (blue squares).

## IV. DESYNCHRONIZATION AND CONTROLLABILITY

The model of hierarchically networked oscillators showed various degrees of synchronization depending on $\lambda_\sigma$ (Fig. 4). This implies that one can control the global synchronization by tuning $\lambda_\sigma$ without changing network topologies or intra- and inter-subsystem coupling strengths ($K$ and $\varepsilon$). Here the adjustment of $\lambda_\sigma$ leads to change the local complexity of subsystems, which then determines the global synchronization between subsystems. For robust realization of desynchronization, Networks 9 and 10 are conspicuous among the 16 networks, because they have the largest parameter region for desynchronization (Fig. 4). Simpler systems with only two oscillators had a very limited parameter region for desynchronization (Fig. S1), which prevented the simple system from producing robust desynchronization.

## V. NUMERICAL METHODS

The differential equations were numerically integrated using the Runge-Kutta 4th order method.[32] We set $r_{n\sigma}(0) = 0.9$ and $\theta_{n\sigma}(0) \in [0, 2\pi]$ as initial conditions, and time step $\Delta t = 0.05$ for simulating a single subsystem ($N$ = 1), and had equilibration period until $t = 10000$. On the other hand, for simulating multiple subsystems ($N$ = 100), we set a smaller time step $\Delta t = 0.01$, and had a longer equilibration period until $t = 40000$. The average degree of synchronization between $N$ = 100 subsystems and the phase relations between three oscillators were obtained from the oscillator phases for the time interval between $t = 40000$ and $t = 50000$. To integrate the differential equations in the presence of noise, we used the Euler method[32] with a small time step $\Delta t = 0.01$, and obtained the inter-subsystem synchronization for the time interval between $t = 4000$ and $t = 5000$. Here, we confirmed that the noise simulation with a smaller time step ($\Delta t = 0.001$) or a longer equilibration period ($t = 40000$) did not change the result significantly.

## VI. DISCUSSION

In this study, we introduced a simple but effective way to predict global synchronization of hierarchical systems by analyzing local complexity of uncoupled subsystems. We considered hierarchical organization of Stuart-Landau oscillators with activity-dependent couplings. Each subsystem consists of three oscillators that interact with each other positively or negatively. Anti-symmetric couplings between three oscillators are special to generate single and multiple dynamic fixed points depending on the amplitudes of the oscillators. Here, the number of attractors can specify the complexity of subsystems. Then, the local complexity can predict the global synchronization of the hierarchical system.

Among possible coupling signs between three oscillators in one subsystem, we found perfect anti-symmetric networks (Networks 9 and 10) are more effective for inducing desynchronization of the whole hierarchical system. However, they can also lead to global synchronization by altering their complexity. The controllability of (de)synchronization may explain why biological networks such as neural networks have asymmetric couplings with excitatory/inhibitory connections. Interestingly, Network 9 is realized in the cellular network of pancreatic islets, critical micro-organs for glucose homeostasis.[28, 31] Each islet (subsystem) consists of three endocrine cells (α, β, and δ cells). Indeed, the synchronization between islets is an important requirement in glucose metabolism.[33] Our finding may suggest the controllability of inter-islet synchronization.[34] It has been recently demonstrated that the unique symmetry of Network 9 for intra-islet interactions can decrease the superficially wasteful zero-sum action of counter-regulatory hormones, and also enhance/suppress the synchronization of hormone secretions between islets under high/normal glucose conditions.[35]

The predictability of global synchronization by local complexity is, however, limited to weakly-coupled hierarchical systems (Fig. S4(b) and Fig. S8), because strong inter-subsystem coupling can generate new attractors and significantly perturb the attractor landscape of uncoupled subsystems. We showed how new fixed points were generated in Network 9 for the case of two subsystems (Fig. 8) with detailed analysis in Appendix. Second, we considered the heterogeneity of intrinsic frequencies $\omega_{n\sigma}$, and confirmed that their large heterogeneity can also diminish the predictability (Fig. 7). Third, we examined noise effects on the predictability. Unless too strong noise makes the system chaotic, the correlation between local complexity and global synchronization is still intact (Fig. 7).

Finally, the predictability can be also sensitive to time delays of coupling, because the time delays can change the effective sign of couplings between oscillators. The change of coupling signs implies different network motifs. Therefore, in this study, we did not investigate the effect of time delays, since the implicit change of coupling signs could be mixed with the explicit differences of coupling signs in network motifs.

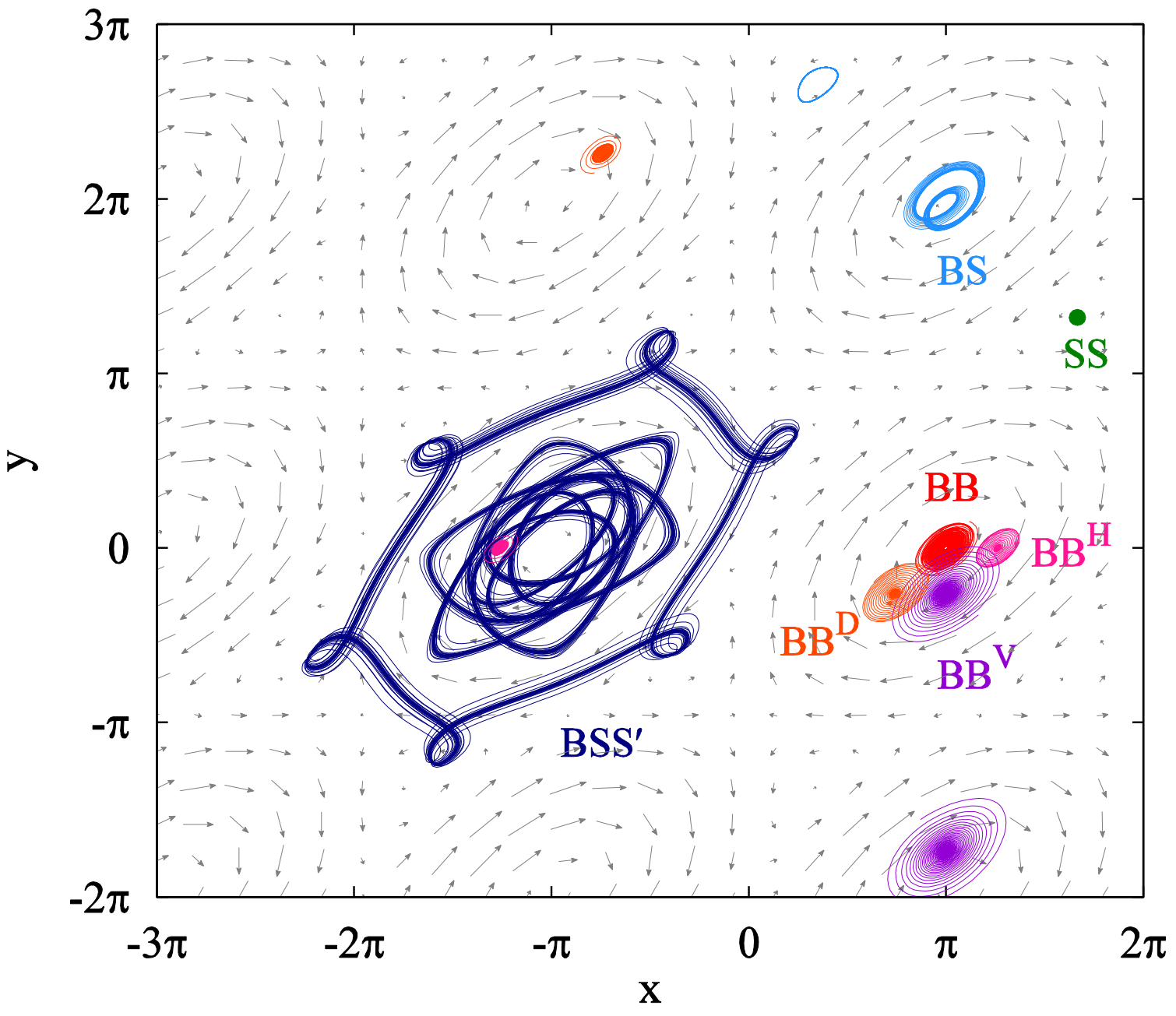

FIG. 8. (Color online) Phase plane of hierarchical systems with two subsystems. Under $\lambda_1 = \lambda_2 = \lambda_3 = 1$, the hierarchical system generates seven dynamic states: (i) *SS* state, in which two subsystems are attracted into the same attractor $S_1$ or $S_2$; (ii) *BS* state, in which one subsystem rotates around $S_1$ or $S_2$, and the other rotates around $B$; (iii) *BSS*' state, in which one subsystem rotates around $B$, and the other subsystem jumps alternately between $S_1$ or $S_2$. (iv) *BB* state, in which two subsystems slowly approach $B$; (v) $BB^H$; (vi) $BB^V$; (vii) $BB^D$ states, in which two subsystems slowly approach horizontally-, vertically-, and diagonally-shifted attractors from $B$. Different colors represent different pairs of two subsystems. For these plots, we used an intra-subsystem coupling strength ($K = 0.1$), identical intrinsic frequency ($\omega_{n\sigma} = 1$) and inter-subsystem coupling strengths ($\varepsilon = 0.1$ or $0.4$) with two subsystems ($N$ = 2). In particular, *BS* state was obtained only with $\varepsilon = 0.1$, while other states were obtained with $\varepsilon = 0.4$. The state realizations depended on initial conditions of phases (Table SI for their specific initial values in Supplementary Material).

**SUPPLEMENTARY MATERIAL**

See Supplementary Material file for Figs. S1-S8 and Table S1 referenced on the text.

## ACKNOWLEDGMENTS

We thank Hyunsuk Hong for helpful comments, and the anonymous reviewers for their valuable comments and suggestions to improve the quality of the paper. This research was supported by Basic Science Research Program through the National Research Foundation of Korea (NRF) funded by the Ministry of Education (2016R1D1A1B03932264) and the Max Planck Society, Gyeongsangbuk-Do and Pohang City (J.J.).

## APPENDIX

The following is the analysis of two-subsystem systems that scrutinizes the synchronization behavior. We first examine the two-subsystem system of Network 9. We again focus on the phase differences ($x_{12}$ and $y_{12}$) between three oscillators within each subsystem, and the phase difference $z$ between two subsystems. The phase variables $(\theta_{11},\theta_{12},\theta_{13},\theta_{21},\theta_{22},\theta_{23})$ of two subsystems are transformed to $(x_1,y_1,x_2,y_2,z)$ where $x_1 \equiv \theta_{11}-\theta_{12}$, $y_1 \equiv \theta_{11}-\theta_{13}$, $x_2 \equiv \theta_{21}-\theta_{22}$, $y_2 \equiv \theta_{21}-\theta_{23}$ and $z \equiv \theta_{11}-\theta_{21}$. Here we assume identical intrinsic frequencies ($\omega_{n\sigma}=\omega$) for simplicity.

First, when $\lambda_\sigma$ are largely dissimilar, individual subsystems have single and identical fixed points (Fig. 3(a)). Thus the phase differences ($x_1, y_1$) and ($x_2, y_2$) for two subsystems are attracted into the same fixed point in the absence of inter-subsystem coupling. The attraction is not perturbed even in the presence of a strong inter-subsystem coupling ($\varepsilon$ = 0.4). The inter-subsystem coupling just contributes to fully synchronize the two subsystems in phase ($\rho$ = 1 and $z$ = 0).

Second, when $\lambda_\sigma$ are similar (e.g. $\lambda_1=\lambda_2=\lambda_3=1$), individual subsystems have three fixed points (Fig. 3(b)), and each subsystem can be attracted to different fixed points ($S_1$, $S_2$, $B$). This complexity impedes the synchronization between the two subsystems. In the absence of inter-subsystem coupling, the two subsystems have four states if one considers the symmetry of $S_1$ and $S_2$: (i) *SS* sate, in which two subsystems stay in the same small basin; (ii) *SS*' state, in which two subsystems stay in different small basins. (iii) *BS* state, in which one subsystem stays in the big basin, while the other subsystem stays in one of the two small basins; and (iv) *BB* state, where two subsystems stay in the big basin. We numerically solved Eqs. (14) and (15) with a strong inter-subsystem coupling ($\varepsilon$ = 0.4), and observed that the two-subsystem system showed modified dynamics in the presence of the coupling (Fig. 8):

(i) *SS* state. Two subsystems arrive at the same attractor, either $S_1$ or $S_2$, and are fully synchronized ($\rho$ = 1).

(ii) *SS*' state. This state is unstable and excluded for $\varepsilon$ = 0.4. Therefore, two subsystems never sit on $S_1$ and $S_2$, respectively.

(iii-1) *BS* state. Two subsystems rotate around *B* and $S_1/S_2$ instead of being attracted into fixed points.

(iii-2) *BSS*' state. One subsystem shows a processional cycle around *B*, and the other subsystem jumps alternately between $S_1$ and $S_2$.

When two subsystems sit on different basins (*BS* and *BSS*' state), they do not show full synchronizations ($\rho$ < 1). Regarding the *BB* state, we found new stationary solutions in the two-subsystem system that satisfy $r_{n\sigma} = \sqrt{\lambda_\sigma} = 1$ with no contribution of coupling terms in Eq. (14):

$$K[\cos x_1 + \cos y_1] + \varepsilon K\left[\cos\left(\bar{x} + \frac{z}{2}\right) + \cos\left(\bar{y} + \frac{z}{2}\right)\right] = 0, \tag{A1}$$

$$K[\cos x_1 - \cos(x_1 - y_1)] + \varepsilon K\left[\cos\left(x_1 - \frac{z}{2}\right) - \cos\left(x_1 - \bar{y} - \frac{z}{2}\right)\right] = 0, \tag{A2}$$

$$K[\cos y_1 + \cos(x_1 - y_1)] + \varepsilon K\left[\cos\left(y_1 - \frac{z}{2}\right) + \cos\left(\bar{x} - y_1 + \frac{z}{2}\right)\right] = 0, \tag{A3}$$

$$K[\cos x_2 + \cos y_2] + \varepsilon K\left[\cos\left(\bar{x} - \frac{z}{2}\right) + \cos\left(\bar{y} - \frac{z}{2}\right)\right] = 0, \tag{A4}$$

$$K[\cos x_2 - \cos(x_2 - y_2)] + \varepsilon K\left[\cos\left(x_2 + \frac{z}{2}\right) - \cos\left(x_2 - \bar{y} + \frac{z}{2}\right)\right] = 0, \tag{A5}$$

$$K[\cos y_2 + \cos(x_2 - y_2)] + \varepsilon K\left[\cos\left(y_2 + \frac{z}{2}\right) + \cos\left(\bar{x} - y_2 - \frac{z}{2}\right)\right] = 0, \tag{A6}$$

With $\bar{x} \equiv (x_1 + x_2)/2$ and $\bar{y} \equiv (y_1 + y_2)/2$. Under these constraints, Eq. (15) yields five phase difference equations:

$$\dot{x}_1 = K[\sin y_1 + \sin(x_1 - y_1)] + \varepsilon K\left[\sin\left(\bar{x} + \frac{z}{2}\right) + \sin\left(\bar{y} + \frac{z}{2}\right) - \sin\left(x_1 - \frac{z}{2}\right) + \sin\left(x_1 - \bar{y} - \frac{z}{2}\right)\right], \tag{A7}$$

$$\dot{y}_1 = K[\sin x_1 + \sin(x_1 - y_1)] + \varepsilon K\left[\sin\left(\bar{x} + \frac{z}{2}\right) + \sin\left(\bar{y} + \frac{z}{2}\right) - \sin\left(y_1 - \frac{z}{2}\right) + \sin\left(\bar{x} - y_1 + \frac{z}{2}\right)\right], \tag{A8}$$

$$\dot{x}_2 = K[\sin y_2 + \sin(x_2 - y_2)] + \varepsilon K\left[\sin\left(\bar{x} - \frac{z}{2}\right) + \sin\left(\bar{y} - \frac{z}{2}\right) - \sin\left(x_2 + \frac{z}{2}\right) + \sin\left(x_2 - \bar{y} + \frac{z}{2}\right)\right], \tag{A9}$$

$$\dot{y}_2 = K[\sin x_2 + \sin(x_2 - y_2)] + \varepsilon K\left[\sin\left(\bar{x} - \frac{z}{2}\right) + \sin\left(\bar{y} - \frac{z}{2}\right) - \sin\left(y_2 + \frac{z}{2}\right) + \sin\left(\bar{x} - y_2 - \frac{z}{2}\right)\right], \tag{A10}$$

$$\dot{z} = K[\sin x_1 + \sin y_1 - \sin x_2 - \sin y_2] + \varepsilon K\left[\sin\left(\bar{x} + \frac{z}{2}\right) - \sin\left(\bar{x} - \frac{z}{2}\right) + \sin\left(\bar{y} + \frac{z}{2}\right) - \sin\left(\bar{y} - \frac{z}{2}\right)\right]. \tag{A11}$$

We examine the stationary conditions, $\dot{x}_1 = \dot{y}_1 = \dot{x}_2 = \dot{y}_2 = \dot{z} = 0$. It is interesting that $\dot{x}_2 = \dot{y}_2 = 0$ is automatically satisfied, once $\dot{x}_1 = \dot{y}_1 = 0$ with constraints, $\bar{x} = (0, \pi)$ and $\bar{y} = (0, \pi)$. The condition gives four *BB* states:

(iv-1) *BB* state ($\bar{x}=\pi$ and $\bar{y}=0$). Two subsystems arrive at *B* ($x_1=x_2=\pi$ and $y_1=y_2=0$) with arbitrary phase difference *z* between two subsystems. Any *z* values can satisfy $\dot{z}=0$ in Eq. (A11); i.e. two subsystems are coupled, but still behave independently.

(iv-2) $BB^H$ state ($\bar{x}=0$ and $\bar{y}=0$). Two subsystems ($x_1$, $y_1$) and ($x_2$, $y_2$) are horizontally located from their center (0, 0); and $y_1=y_2=0$ and $x_1(=-x_2)$ and *z* satisfy

$$e^{ix_1}+2\varepsilon e^{iz/2}=-1. \quad \text{(A12)}$$

(iv-3) $BB^V$ state ($\bar{x}=\pi$ and $\bar{y}=\pi$). Two subsystems ($x_1$, $y_1$) and ($x_2$, $y_2$) are vertically located from their center (π, π); $x_1=x_2=\pi$ and $y_1(=2\pi-y_2)$ and *z* satisfy

$$e^{iy_1}-2\varepsilon e^{iz/2}=1. \quad \text{(A13)}$$

(iv) $BB^D$ state ($\bar{x}=0$ and $\bar{y}=\pi$). Two subsystems ($x_1$, $y_1$) and ($x_2$, $y_2$) are diagonally located from their center (0, π); $x_1-y_1=-\pi$ and $y_1(=2\pi-y_2)$ and *z* satisfy

$$e^{iy_1}+2\varepsilon e^{i(y_1-z/2)}=1. \quad \text{(A14)}$$

The complex Eqs. (A13)-(A14) give exact values of the phase differences ($x_1$, $y_1$) and ($x_2$, $y_2$) between three oscillators within each subsystem, and the phase difference *z* between two subsystems. Therefore, $BB^H$, $BB^V$, *and* $BB^D$ states can emerge in the presence of inter-subsystem coupling ($\varepsilon\neq 0$). They prefer out-of-phase coordination between two subsystems ($z\approx 3.34$, $\rho\approx 0.1$) for a weak coupling ($\varepsilon=0.1$), while they prefer in-phase coordination ($z\approx 5.38$, $\rho\approx 0.9$) for a strong coupling ($\varepsilon=0.9$).

TABLE I. State probability of two-subsystem systems.

| X | $P(X\mid\varepsilon=0)$ | $P(X\mid\varepsilon=0.4)$ |
|---|---|---|
| *SS* | 0.057±0.001 | 0.078±0.003 |
| *SS*’ | 0.057±0.003 | 0 |
| *BS* | 0.449±0.005 | 0 |
| *BSS*’ | 0 | 0.258±0.005 |
| *BB* | 0.437±0.004 | 0.151±0.005 |
| $BB^H$ | 0 | 0.208±0.005 |
| $BB^V$ | 0 | 0.154±0.004 |
| $BB^D$ | 0 | 0.152±0.004 |

The realization of these seven states depends on initial conditions, $[r_{n\sigma}(0),\theta_{n\sigma}(0)]$. Since the two-subsystem system has *riddled basins* of which attractors are densely mixed up, it is difficult to predict toward which attractors an arbitrary initial condition will flow to. Although the prediction is uncertain, the realization probability of each state is not uncertain (Table I). The probabilities of eight states are obtained for the two-subsystem systems in the absence ($\varepsilon=0$) and presence ( $\varepsilon = 0.4$ ) of inter-subsystem coupling, given $\lambda_1 = \lambda_2 = \lambda_3 = 1$, $K$ = 0.1, and $\omega_{n\sigma} = 1$. We ran $10^4$ different initial conditions to estimate the state probability, and computed the standard deviation from 10 sets of ensembles to obtain their uncertainties. The frequency of *SS* state, which generates full synchronization ($\rho$ = 1), is relatively small because two subsystems rarely enter the same small basin. Therefore, as the number of subsystems considered increases, the likelihood that each subsystem sits on a different attractor also increases, so the subsystems in a hierarchical system may become desynchronized.

We performed the same analysis for Network 10, and confirmed that the two-subsystem system of Network 10 also generate seven states like Network 9. Their realization probabilities are also the same except for the switch between *BB* and $BB^H$ states (Table I): $P(BB \mid Network10) = P(BB^H \mid Network9)$ and $P(BB^H \mid Network10) = P(BB \mid Network9)$. Summarizing the analysis of the two-subsystem systems, the two anti-symmetric subsystems generate new fixed points in the presence of inter-subsystem coupling ( $\varepsilon \neq 0$ ). Then their realization probabilities are different between Network 9 and Network 10, although they show the same complexity map in the absence of inter-subsystem coupling ( $\varepsilon = 0$ ).

Supplementary Material

# Local complexity predicts global synchronization of hierarchically networked oscillators

Jin Xu,[1,2,a)] Dong-Ho Park,[1,a)] and Junghyo Jo[1,2,b)]

[1]*Asia Pacific Center for Theoretical Physics (APCTP), 67 Cheongam-ro, Pohang, 37673, Korea*

[2]*Department of Physics, Pohang University of Science and Technology (POSTECH), Pohang, 37673, Korea*

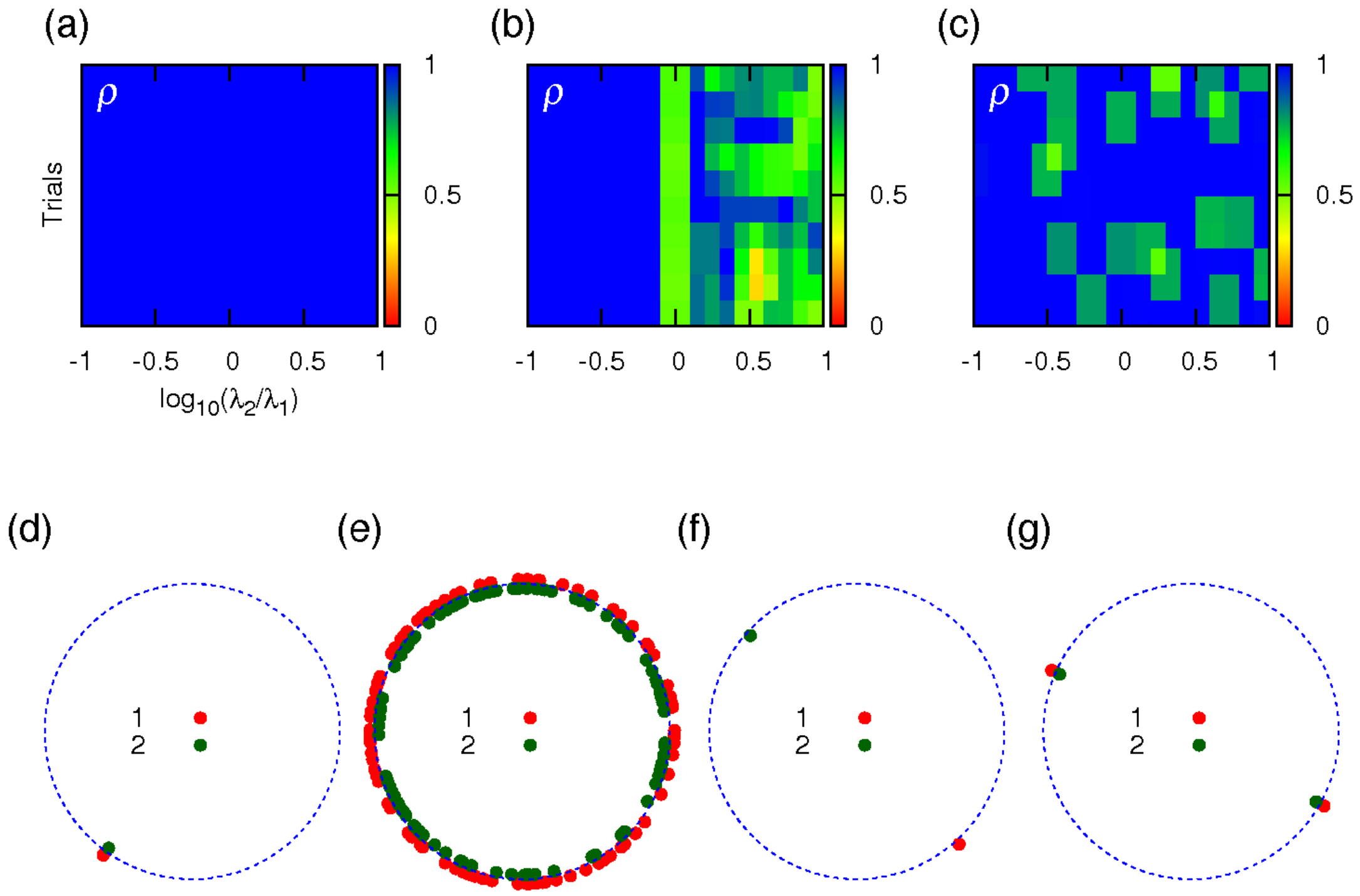

FIG. S1. Global synchronization and phase dynamics of subsystems consisting of two oscillators. Three possible interaction symmetries were considered: (a) mutual activation, (b) asymmetric interaction, and (c) mutual inhibition. Depending on the values of two activity parameters ( $\lambda_1$ and $\lambda_2$ ), average degree of synchronization ( $\rho$ ) between subsystems and phase relations between two oscillators in different subsystems were obtained after equilibration. To show initial condition dependence, we plotted ten trials and x-axis with 21 bins. For mutual activation, every oscillator is synchronized in phase (d). For mutual inhibition, two populations of oscillators are out of phase, sometimes each population was fully synchronized (f) or two clusters of each oscillator was synchronized while they are keeping out of phase (g) (In this case, they show partial synchronization in (c)). Finally, the asymmetric interaction shows both in-phase synchronization (d) in the activation dominant regime ( $\lambda_1 > \lambda_2$ ) and out-of-phase synchronization (f) and (g) in the inhibition dominant regime ( $\lambda_2 > \lambda_1$ ). And for $\lambda_1 = \lambda_2$ , two populations of oscillators are partially synchronized (e). For these plots, we used an intra-subsystem coupling ( $K = 0.1$ ), and identical intrinsic frequency ( $\omega_{n\sigma} = 1$ ).

---

a) These authors are equally contributed.
b) Author to whom correspondence. Electronic mail: jojunghyo@apctp.org

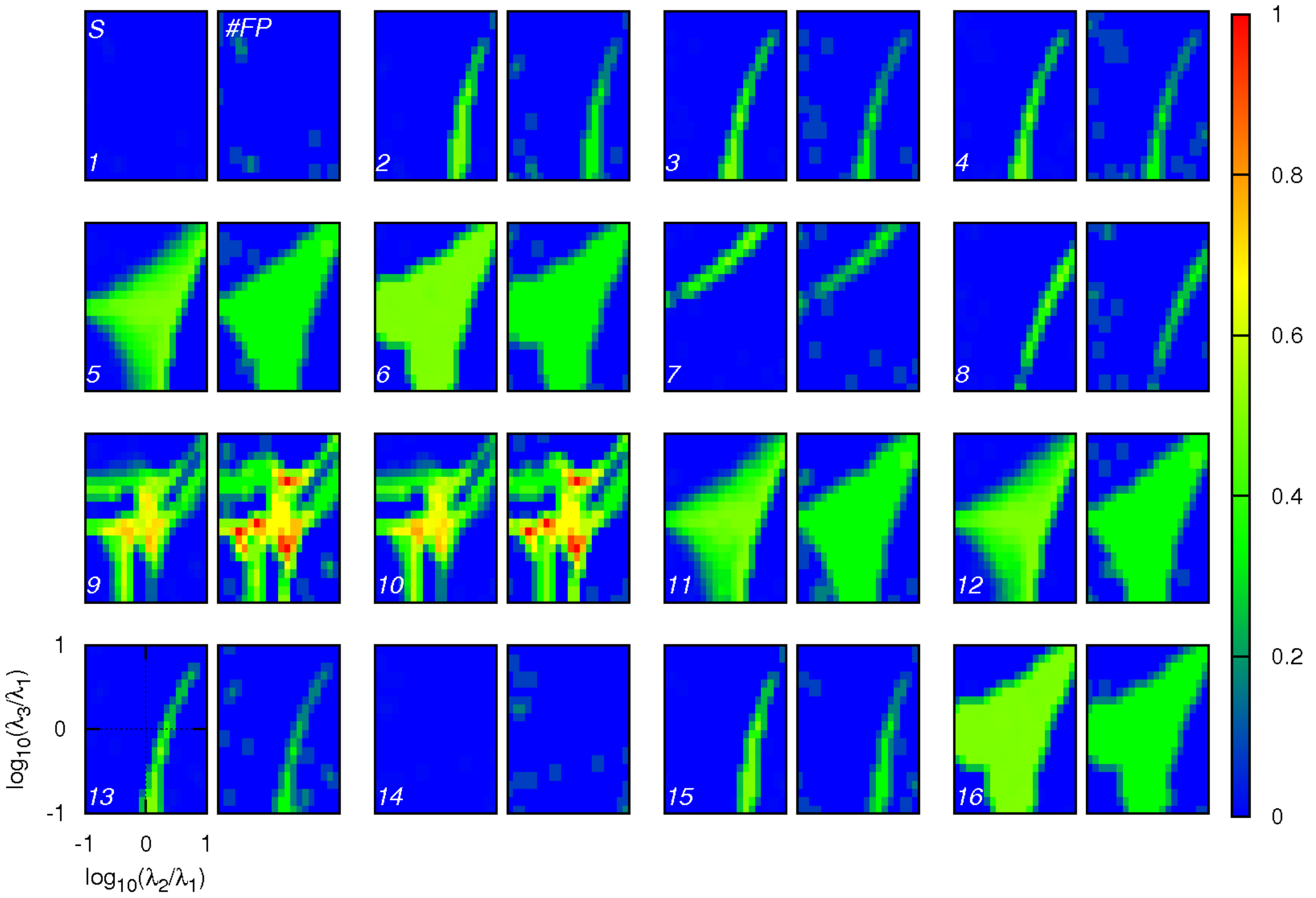

FIG. S2. Number of fixed points (#FP) on the phase plane of the three coupled oscillators within a subsystem and the local subsystem complexity ($S$). The subsystem complexity ($S$, left panel) is measured by the Shannon entropy of realized fixed points ($x^*$, $y^*$) of three coupled oscillators within a certain accuracy ( $\delta x^*, \delta y^* < \pi / 5$ ). One hundred ensembles are used to obtain the frequencies of realized fixed points. For clear comparison, we normalized the two variables by $S$/2 and (#FP-1)/3. Axes have base-10-logarithmic scales to allow presentation of a large range of $\lambda_\sigma$ , and each axis has 21 bins. For these plots, we used an intra-subsystem coupling ( $K = 0.1$ ), and identical intrinsic frequency ( $\omega_{n\sigma} = 1$ ).

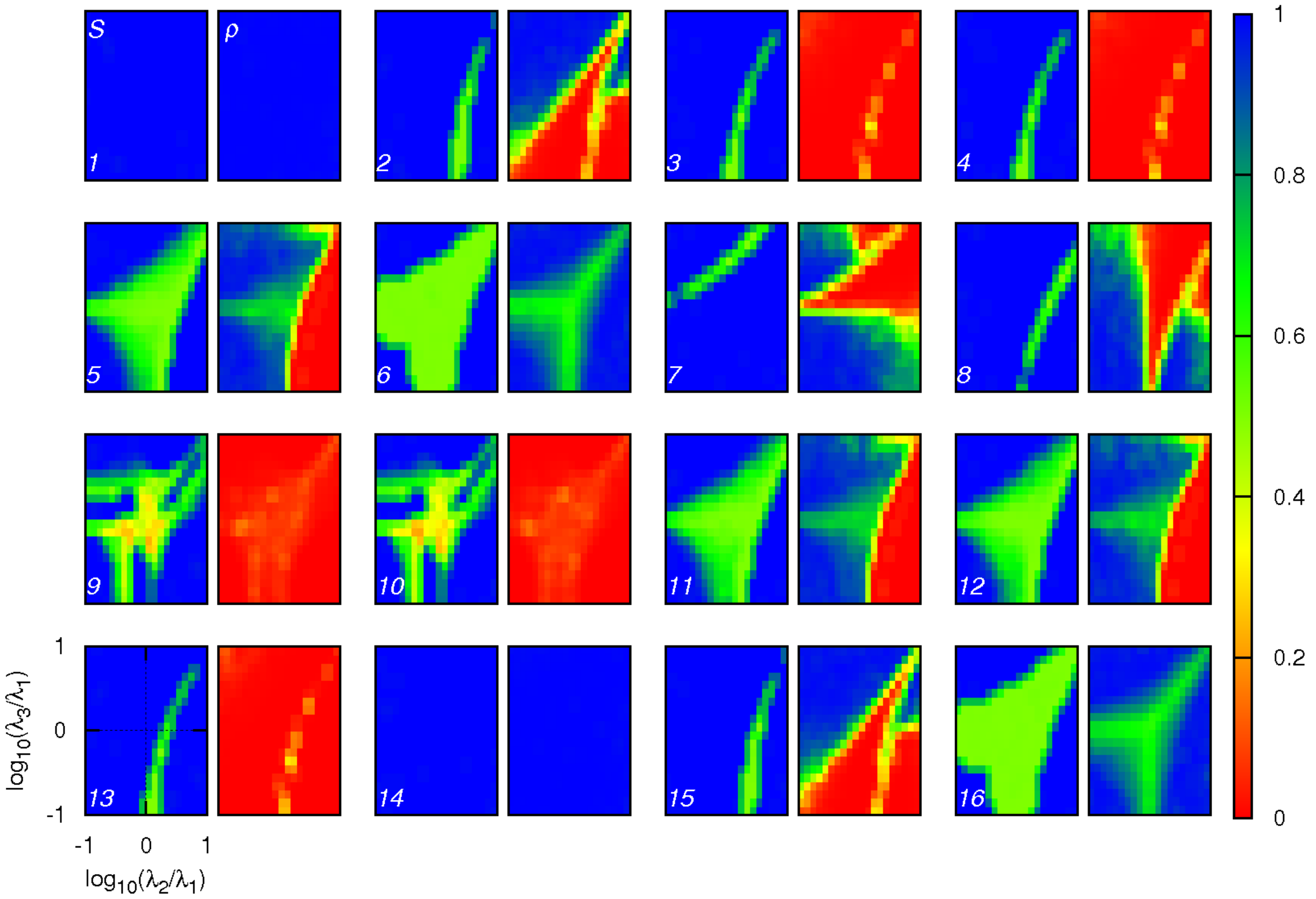

FIG. S3. Local complexity and global synchronization for the inter-subsystem coupling by arithmetic mean fields. The subsystem complexity ($S$, left panel) is measured by the Shannon entropy of realized fixed points ($x$*, $y$*) of three coupled oscillators within a certain accuracy ( $\delta x^*, \delta y^* < \pi/5$ ). One hundred ensembles are used to obtain the frequencies of realized fixed points. The average degree of synchronization ($\rho$, right panel) between subsystems is measured. For clear comparison with $\rho$ (red: desynchronization; blue: synchronization), the complexity index $S$ is reduced by half (red: 1; blue: 0). Axes have base-10-logarithmic scales to allow presentation of a large range of $\lambda_\sigma$ , and each axis has 21 bins. For these plots, we used an intra-subsystem coupling ( $K = 0.1$ ), an inter-subsystem coupling ( $\varepsilon = 0.1$ ), and identical intrinsic frequency ( $\omega_{n\sigma} = 1$ ) for $N$ = 100 subsystems.

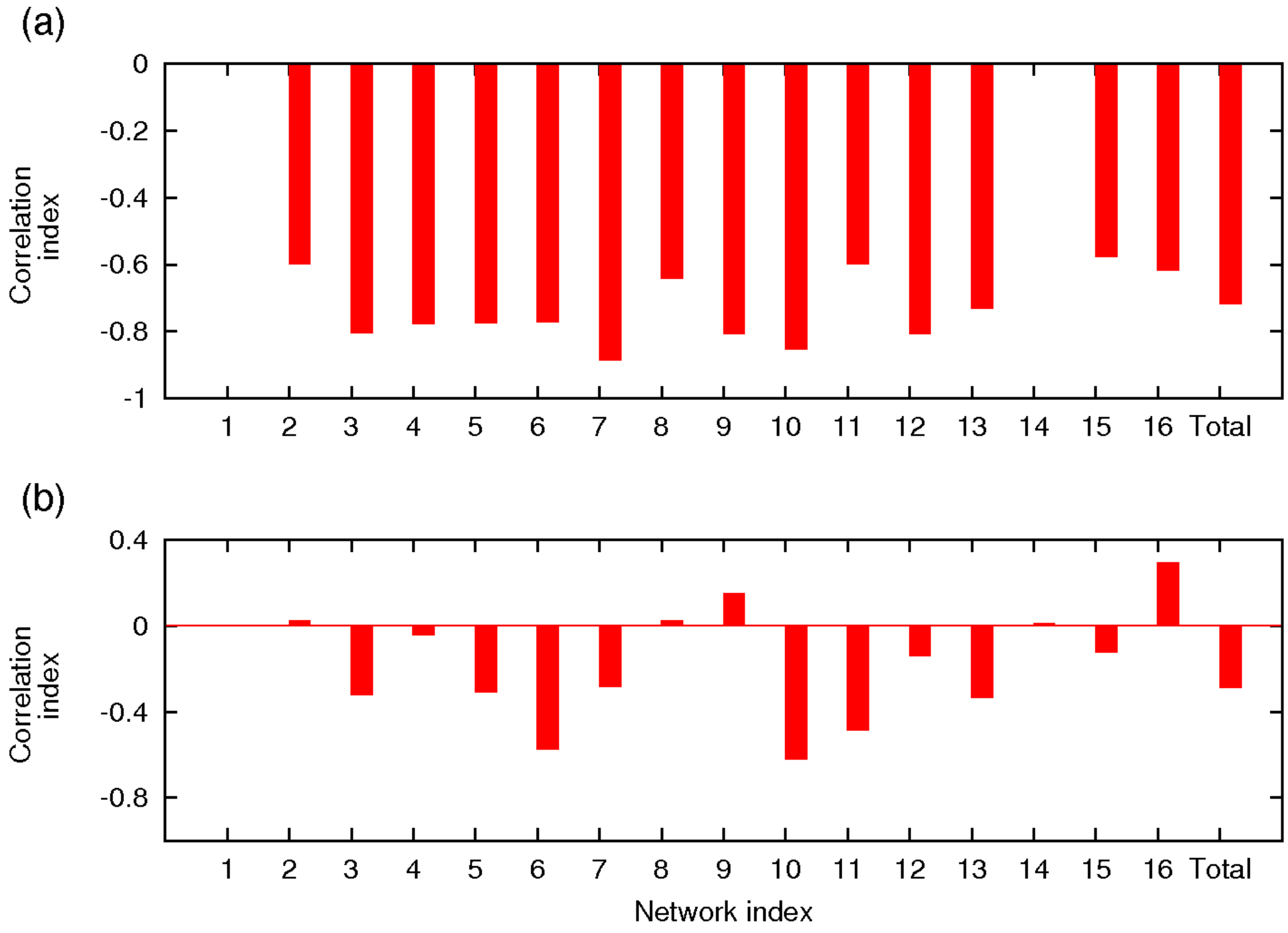

FIG. S4. Pearson correlations between local complexity and global synchronization for the 16 networks. For the simulation of the hierarchical system, we used inter-subsystem coupling strengths (a) $\varepsilon = 0.1$ and (b) $\varepsilon = 0.4$. The correlations were absent in Network 1 and 14 (on (a)), because they showed perfect synchronization ($\rho$ = 1) independent of the local complexity.

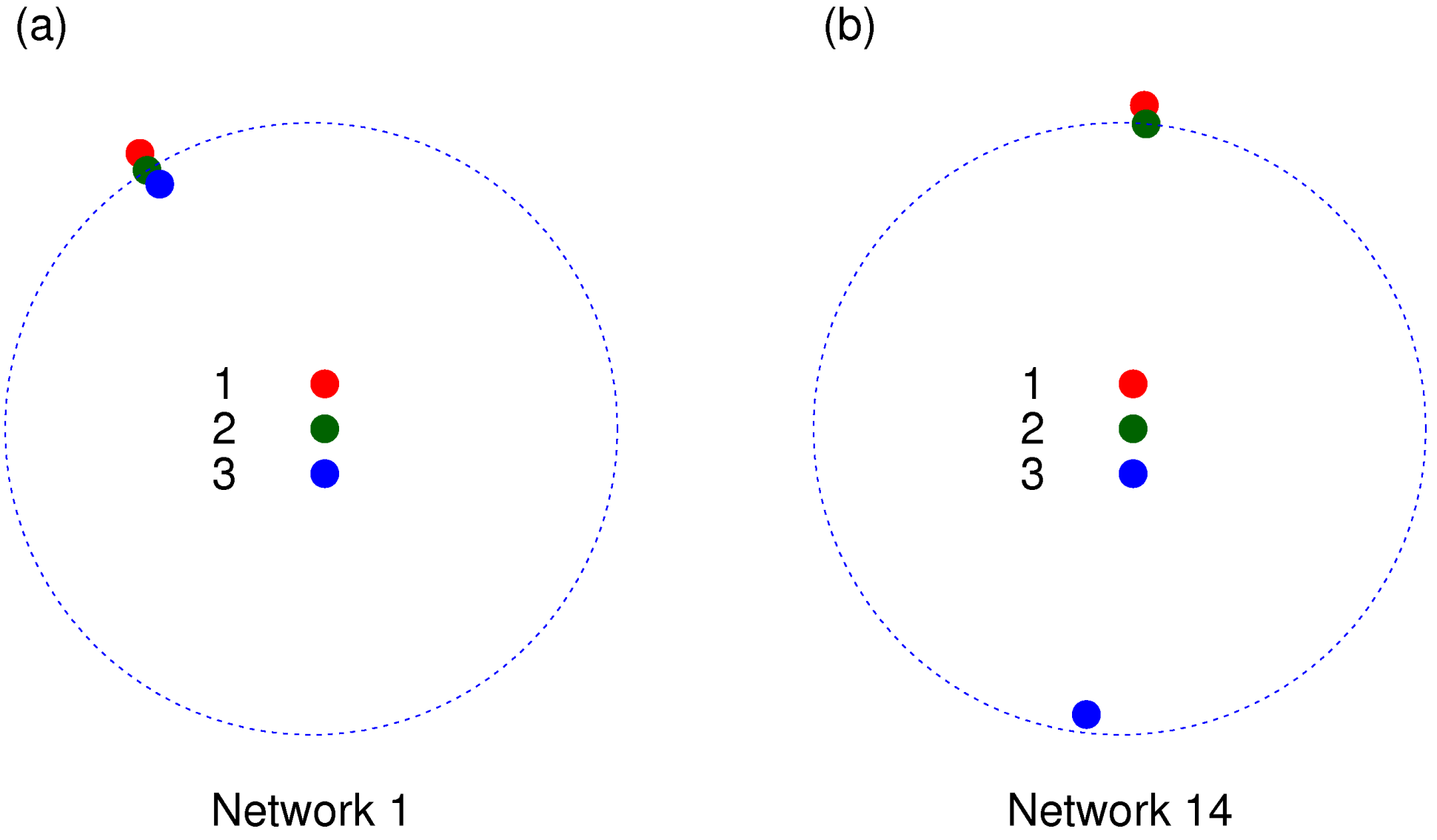

FIG. S5. One phase state for Networks 1 and 14. After equilibration of phase dynamics, for $\lambda_1 = \lambda_2 = \lambda_3 = 1$ , $\theta_{n1}$ , $\theta_{n2}$ , and $\theta_{n3}$ are all synchronized in phase for Network 1; $\theta_{n1}$ and $\theta_{n2}$ are synchronized in phase, but $\theta_{n1}$ and $\theta_{n3}$ are out of phase for Network 14. $n \in \{1,...,100\}$.

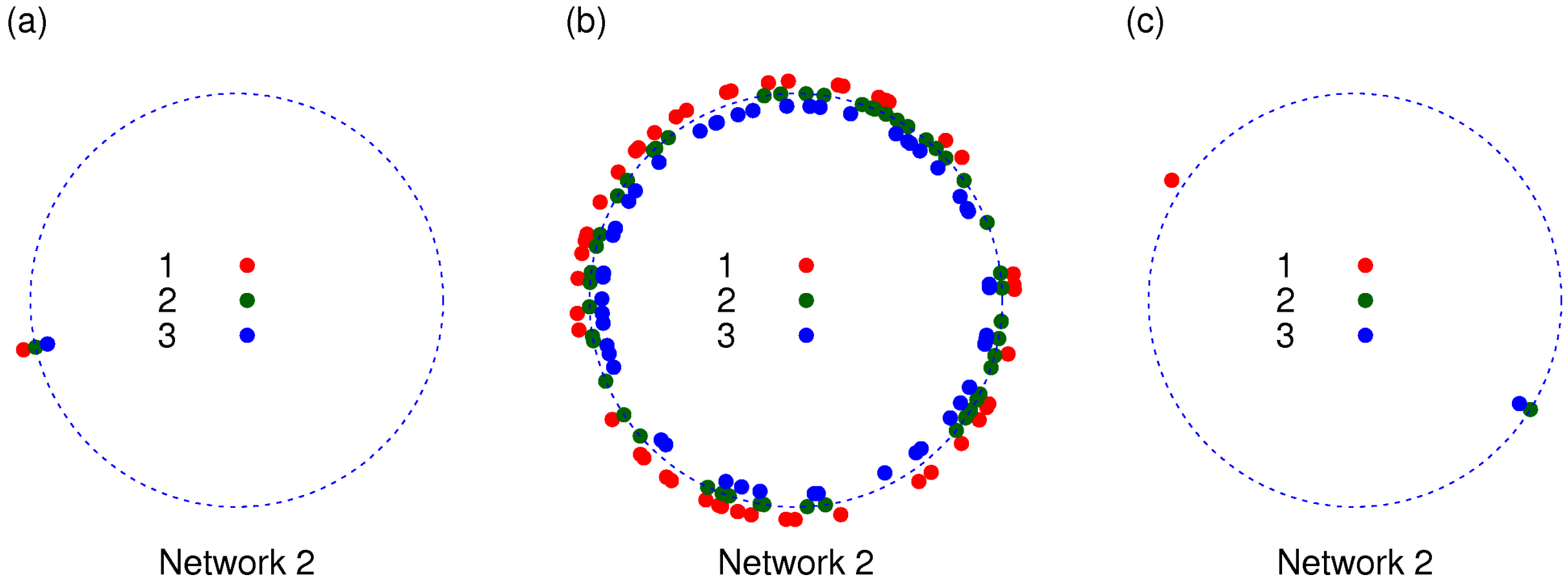

FIG. S6. Phase dynamics of two phase states for Network 2. After equilibration, the phase states of Network 2 depend on parameters $\lambda_\sigma$ . (a) For $\lambda_1 = 1.0$ , $\lambda_2 = 1.0$ and $\lambda_3 = 10.0$ , $\theta_{n1}$ , $\theta_{n2}$ , and $\theta_{n3}$ are synchronized in phase. (b) For the transition region ( $\lambda_1 = 1.0$ , $\lambda_2 = 2.0$ and $\lambda_3 = 0.1$ ), oscillators are desynchronized.(c) For $\lambda_1 = 1.0$ , $\lambda_2 = 10.0$ and $\lambda_3 = 0.1$ , $\theta_{n1}$ and $\theta_{n2}$ are synchronized out of phase, while $\theta_{n2}$ and $\theta_{n3}$ are synchronized in phase.

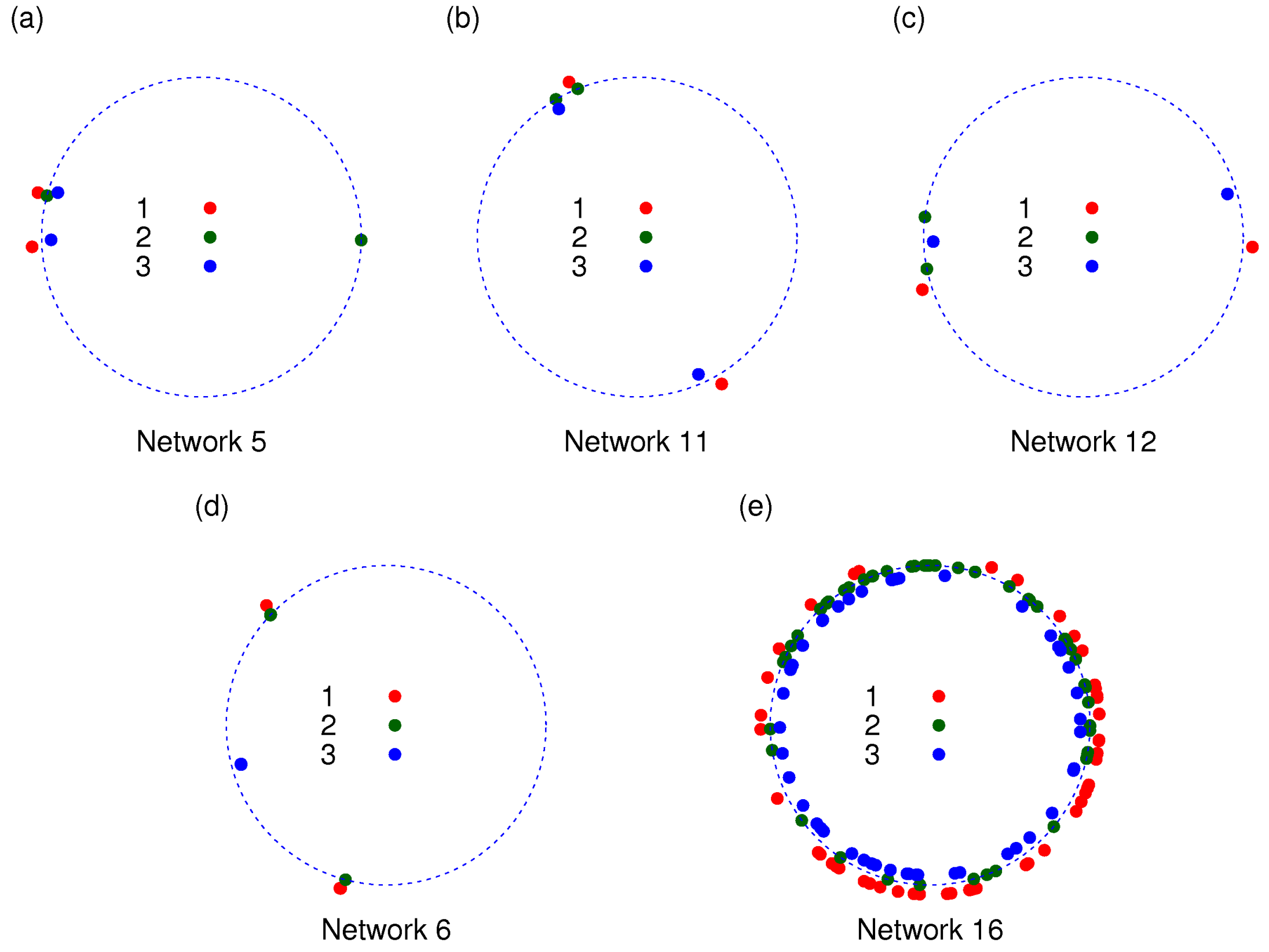

FIG. S7. Phase dynamics of three phase states for Networks 5, 6, 11, 12, and 16. After equilibration of phase dynamics for $\lambda_1 = \lambda_2 = \lambda_3 = 1$ , Networks 5, 11, and 12 showed two synchronized clusters of oscillators 1, 2, and 3, while Network 6 showed two synchronized clusters of oscillators 1, 2, and one synchronized cluster of oscillators 1. However, Network 16 did not showed the synchronized clusters, but showed partial synchronization between oscillators.

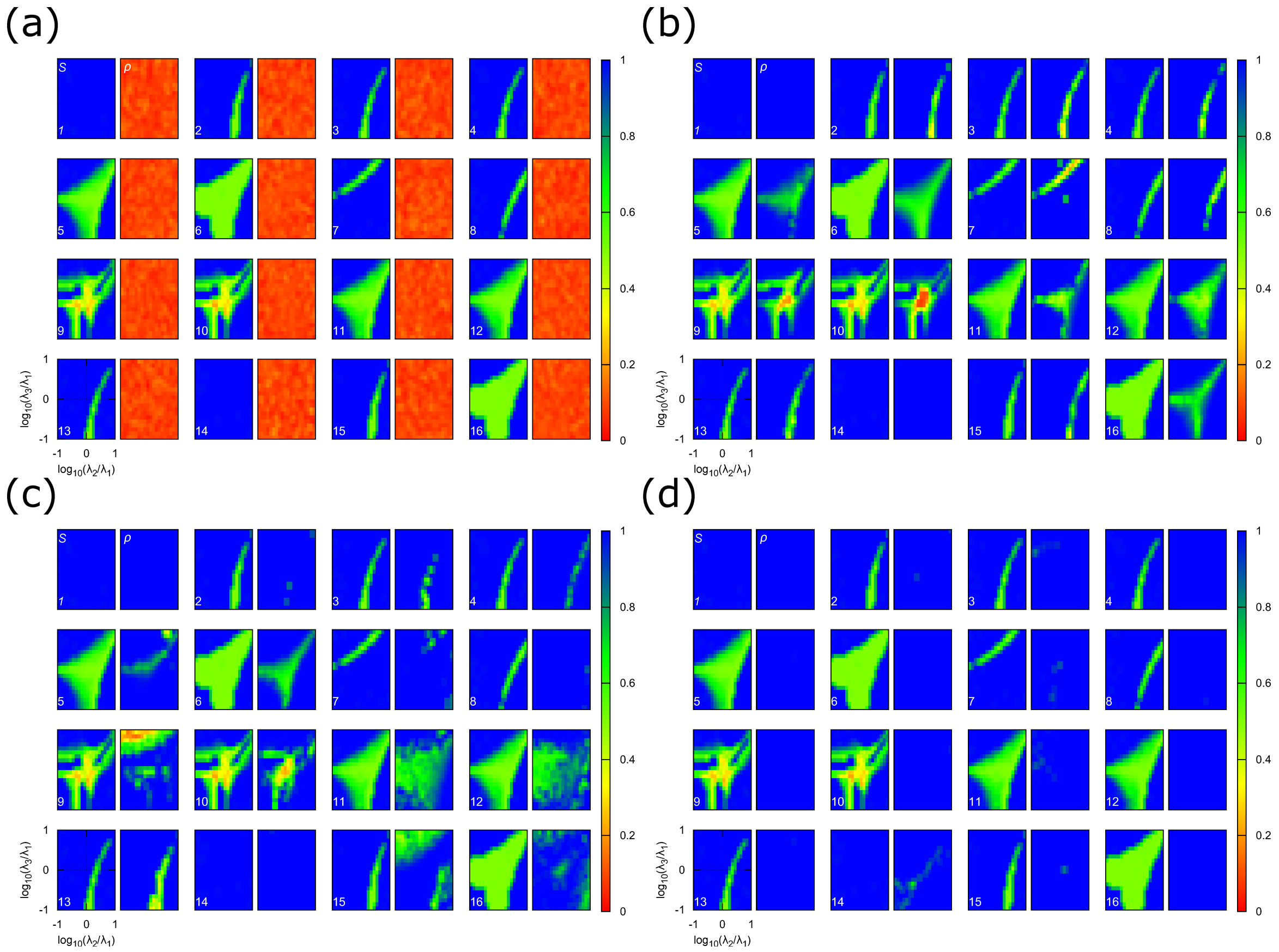

FIG. S8. Local complexity and global synchronization for different inter-subsystem coupling strengths. The subsystem complexity ($S$, left panel) is measured by the Shannon entropy of realized fixed points ($x^*$, $y^*$) of three coupled oscillators within a certain accuracy ($\delta x^*, \delta y^* < \pi/5$). One hundred ensembles are used to obtain the frequencies of realized fixed points. The average degree of synchronization ($\rho$, right panel) between 100 subsystems is measured. For clear comparison with $\rho$ (red: desynchronization; blue: synchronization), the complexity index $S$ is reduced by half (red: 1; blue: 0). Axes have base-10-logarithmic scales to allow presentation of a large range of $\lambda_\sigma$. For the simulation of the hierarchical system, we used $\varepsilon =$ (a) 0, (b) 0.1, (c) 0.4, and (d) 5.0 (inter-subsystem coupling) between. Axes have base-10-logarithmic scales to allow presentation of a large range of $\lambda_\sigma$, and each axis has 21 bins. For these plots, we used an intra-subsystem coupling ($K = 0.1$), and identical intrinsic frequency ($\omega_{n\sigma} = 1$) for $N = 100$ subsystems.

TABLE SI. Initial conditions for various states of two coupled subsystems.

| States | $\theta_{11}$ | $\theta_{12}$ | $\theta_{13}$ | $\theta_{21}$ | $\theta_{22}$ | $\theta_{23}$ |
|---|---|---|---|---|---|---|
| *SS* | 0.0 | -0.1 | -1.6 | 0.0 | -0.5 | -2.6 |
| *BS* | 0.0 | -0.1 | -1.6 | 0.0 | -2.1 | 1.6 |
| *BSS'* | 1.6 | 0.6 | 0.7 | 3.3 | 1.4 | 1.5 |
| *BB* | 4.0 | 0.6 | 6.1 | 4.2 | 1.2 | 2.7 |
| $BB^H$ | 0.0 | -0.1 | -1.6 | 0.0 | -5.1 | 1.6 |
| $BB^V$ | 0.0 | -0.1 | -1.6 | 0.0 | -2.5 | -2.5 |
| $BB^D$ | 5.1 | 6.2 | 0.6 | 6.0 | 3.2 | 5.6 |

Note that the phase units are radian, and initial amplitudes were always set as $r_{n\sigma}(0) = 0.9$ .